# Exploring the Growth Dynamics of Size-selected Carbon Atomic Wires with *in situ* UV Resonance Raman Spectroscopy

Pietro Marabotti[a,b]*, Sonia Peggiani[a], Simone Melesi[a], Barbara Rossi[c], Alessandro Gessini[c], Andrea Li Bassi[a], Valeria Russo[a], Carlo Spartaco Casari[a]*

[a] Department of Energy, Micro and Nanostructured Materials Laboratory - NanoLab, Politecnico di Milano, Via Ponzio 34/3, Milano 20133, Italy
[b] Institut für Physik, Humboldt Universität zu Berlin, 12489 Berlin, Germany
[c] Elettra Sincrotrone Trieste, S.S. 114 km 163.5, Basovizza, 34149 Trieste, Italy.

*Corresponding authors: pietro.marabotti@polimi.it, carlo.casari@polimi.it

# Abstract

Short carbon atomic wires, the prototypes of the lacking carbon allotrope carbyne, represent the fundamental one-dimensional system and the first stage in carbon nanostructure growth, which still exhibits many open points regarding their growth and stability. We introduce an *in situ* UV resonance Raman approach for real-time monitoring of the growth of carbon atomic wires during pulsed laser ablation in liquid without perturbing the synthesis environment. We track single-chain species' growth dynamics, achieving size selectivity by exploiting the peculiar optoelectronic properties of carbon wires and the tunability of synchrotron radiation. We systematically explore diverse solvents, finding size- and solvent-dependent production rates linked to the solvent's C/H ratio and carbonization tendency. Carbon atomic wires' growth dynamics reveal a complex interplay between formation and degradation, leading to an equilibrium. Water, lacking in carbon atoms and reduced polyynes' solubility, yields fewer wires with rapid saturation. Organic solvents exhibit enhanced productivity and near-linear growth, attributed to additional carbon from solvent dissociation and low relative polarity. Exploring the dynamics of the saturation regime provides new insights into advancing carbon atomic wires' synthesis via PLAL. Understanding carbon atomic wires' growth dynamics can contribute to optimizing PLAL processes for nanomaterial synthesis.

# Introduction

Carbon's remarkable ability to assemble in diverse structures has captivated researchers' interest across many scientific and technological fields, ranging from astronomy to material science and nanotechnology. In astrochemistry, fullerene was first detected in space [1], and the presence of carbon clusters in interstellar dust is pivotal to understanding the formation of large organic molecules, eventually related to the origin of life [2,3]. Short linear sp-carbon chains or carbon atomic wires have been detected in diffuse interstellar bands and carbon clusters synthesized in a laboratory. Further, they are considered a key step in the formation of larger carbon aggregates, including fullerenes (*e.g.*, the so-called fullerene road) as the first stages of growth under strong non-equilibrium conditions, possibly explaining their elusive nature [4–10].

Carbon atomic wires are among the most intriguing materials currently under investigation. These linear systems are the finite realization of carbyne, the ideal one-dimensional linear chain made of sp-hybridized carbon atoms [11–13]. The interest in these systems arises from the outstanding properties predicted for carbyne, ranging from superior optical absorption and thermal transport to the transition between a metallic phase (*i.e.*, cumulene) and a semiconducting one (*i.e.*, polyyne) [14]. Confined carbyne, *i.e.*, long (>1000 carbon atoms) linear carbon chains encapsulated in double-walled carbon nanotubes, represents the closest approach to carbyne [12,15]. Researchers recently discovered its exceptional optical properties, like its paramount resonance Raman cross section, higher than any other known material [14,16–18]. In this framework, carbon atomic wires represent small carbyne-like chains with less than 100 sp-hybridized carbon atoms and end-capped at both edges by a large set of possible functional groups [11–13]. It is possible to tailor carbon atomic wires' optical, thermal, mechanical, and electronic properties by varying their length and termination [19–32]. However, this requires high control and flexibility in the synthesis process. Chemical synthesis methods provide an appropriate control, producing chains with selected lengths and terminations, but they are not flexible and fast enough to allow a proper scale-up to an industrial production level [11,20,33]. On the other hand, physical synthesis techniques possess these capabilities, but they grant only limited control over the synthesis process and final products so far [11,34,35].

Among the physical techniques, pulsed laser ablation in liquid (PLAL) is the most versatile, flexible, and simple method to synthesize a wide range of carbon atomic wires [36]. It employs a short laser pulse, from the *fs* to the *ns* range, to irradiate a solid target or a powder immersed in a liquid medium or focus the pulse within the solvent. This technique allows to obtain chains with different sizes and terminations by selecting the solvent [37–50], target [34,38,51–57], pulse duration [42,58–64], laser wavelength [34,51,65,66], and energy of the laser beam [34,36,42] properly. Even if many works report carbon atomic wires' synthesis *via* PLAL, many aspects governing the formation process remain elusive [36,51]. The linear chains' growth is intricately influenced by the interplay between two competing phenomena: polymerization reactions, driving chain elongation, and hydrogenation reactions, which result in chain termination, typically with hydrogen atoms [36,51]. This synthesis process is inherently based on the interaction of radical species, and the polymerization is believed to occur through adding carbon dimers and/or ethynyl radicals [36,48,51]. Unfortunately, direct proof of these processes during ablation remains challenging due to their ultrafast timescales (ranging from *fs* to hundreds of *ns* [36,67–69]) and the complex environment of PLAL experiments. In this context, *in situ* (*i.e.*, during the growth) experiments hold promise for deepening our understanding even though a characterization technique with high sensitivity to the local chemical bond is required to monitor the formation process.

Raman spectroscopy provides a powerful characterization technique for carbon atomic wires and a non-disruptive, contactless, and fast method to monitor them and eventually develop an *in situ* diagnostic tool. Indeed, carbon atomic wires possess a remarkable Raman response, which consists of a fingerprinting Raman-active vibration called ECC (from the Effective Conjugation Coordinate theory) or α mode [70,71]. The α mode ranges from 1800 to 2300 cm$^{-1}$, in a spectral region where other carbon allotropes, like byproducts simultaneously produced with carbon atomic wires during the ablation, do not exhibit any Raman-active mode [71]. Its frequency modulates with the chain length and termination, providing a unique Raman signal for each species and allowing us to track single-chain species dynamics [71]. Depending on the chain length and terminations, other Raman-active vibrations are present in the characteristic frequency range of the α mode, like, for example, the β mode of hydrogen-capped polyynes [25,70,72] or the CN stretching mode of cyano-capped polyynes [72].

Unfortunately, the typical concentration of carbon atomic wires in mixtures produced by PLAL is well below the threshold (≈10$^{-3}$ mol/L) to collect good-quality Raman spectra, and some signal enhancer is required. In a previous work, we conducted an *in situ* experimental campaign using a surface-enhanced Raman scattering (SERS) probe [73]. Our findings revealed an intriguing aspect: carbon atomic wires degrade already during their synthesis, influencing the final yield and chains' size distribution. However, our SERS data exhibit the convolution of SERS signals from wires of different lengths and terminations and are mediated by the interaction between carbon atomic wires and metal nanoparticles. Hence, gaining insights into the growth dynamics of size- and termination-selected wires remains challenging.

We recently showed how resonance Raman spectroscopy resulted in an outstanding characterization method for carbon atomic wires and, in particular, polyynes. Indeed, by finely tuning the excitation wavelengths to the electronic transitions of carbon atomic wires, their Raman signal can be significantly enhanced, making low concentrated samples, down to 10$^{-8}$ mol/L, observable [25,72]. Moreover, polyynes possess intense and sharp electronic transition in the UV range (from 198 to 400 nm), whose energies modulate with the chain length and terminations [11,37,45,74], providing the required selectivity to follow the dynamics of the single chain in a polydisperse mixture, *i.e.*, observing the resonance Raman signal of wires selected by length and termination in a polydispersed solution obtained through PLAL. The increment of the detection limit due to the resonance enhancement allows us to collect Raman spectra in a reasonable time interval (*i.e.*, a few seconds) compared to the characteristic ablation times (*i.e.*, several minutes). Based on these results, resonance Raman spectroscopy is the candidate technique for real time *in situ* monitoring of the growth of carbon atomic wires during their synthesis by laser ablation in liquid with size-selected resolution.

In this work, we designed and set up an *in situ* multi-wavelength UV resonance Raman system to monitor the *real time* growth of carbon atomic wires through our PLAL apparatus. The fine tunability required to match the resonance condition with each different wire (*i.e.*, step of ≈1 nm in the 200-272 nm range) is provided using the synchrotron radiation as Raman excitation source. We perform ablation in water and organic solvents (methanol, isopropanol, and acetonitrile), selected for their UV transparency, stability under laser ablation, and varying polarity and C/H ratio. This selection allows us to explore how these key parameters influence polyyne formation yield and stability. Instead, water was included as a negative control to highlight the role of organic solvents in providing additional carbon sources. We track the behavior of four size-selected hydrogen-capped polyynes, *i.e.*, $HC_8H$, $HC_{10}H$, $HC_{12}H$, and $HC_{14}H$. We develop an effective, simple, and analytical method to correct the raw Raman data from self-absorption induced by the remarkable absorption of the ablation mixture at the Raman excitation wavelength and extract the concentration of each wire as a function of the ablation time. We detect the onset of degradation processes simultaneously with the formation of carbon atomic wires, eventually leading to an equilibrium between formation and degradation. Observing the dynamics of carbon atomic wires' growth, we evaluate the production rate during ablation and other relevant parameters, like the maximum concentration reachable for each chain (*i.e.*, at the formation-degradation equilibrium) and the time needed to reach that concentration. Our findings provide further insight into carbon atomic wires' growth dynamics and carbon nanostructures in strong out-of-equilibrium conditions and the effect of the liquid environment on the growth of size-selected chains.

## Results and Discussion

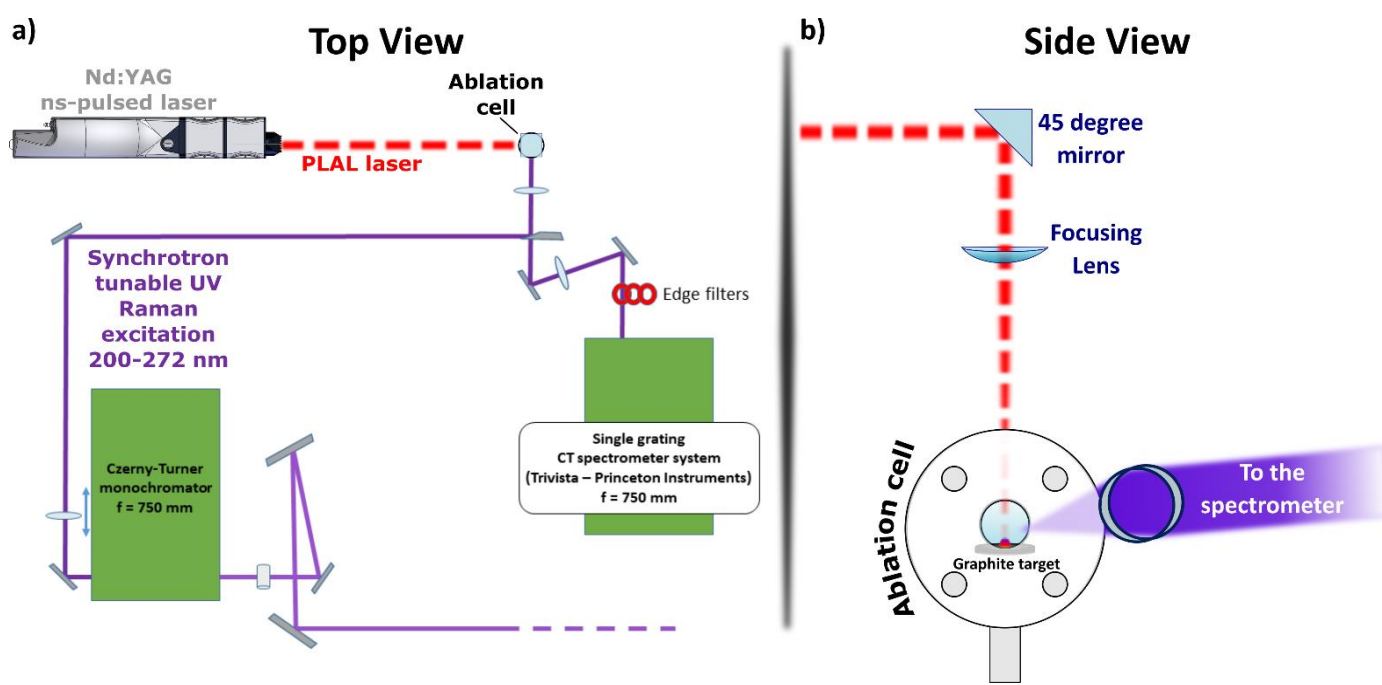

**Figure 1 Pulsed laser ablation setup integrated with UV Raman system.** a) Top and b) side views of the pulsed laser ablation setup integrated with the synchrotron-based UV resonance Raman system of the IUVS beamline to perform *in situ* monitoring of the growth of polyynes.

We *in situ* monitored the growth of carbon atomic wires (*i.e.*, polyynes) during pulsed laser ablation in liquid (PLAL) experiments by integrating our PLAL system with the multi-wavelength, synchrotron-based UV Raman scattering setup of the IUVS beamline (Elettra synchrotron, Italy), as illustrated in Figure 1. After optimizing the signal, we set an acquisition time of 10 s per spectrum, which is fast enough to provide reliable statistics for the whole growth dynamics (15 minutes) and adequately long to obtain a good signal-to-noise ratio during the early stages of ablation.

The exceptional selectivity of UV resonance Raman (UVRR) allowed us to track the evolution of single-chain species. As a proof of concept, we collected two UVRR spectra of polydispersed mixtures containing different carbon atomic wires and byproducts. After tuning the synchrotron radiation to a selected wavelength in resonance with the most intense vibronic transition of a specific polyyne, the UVRR spectra of these mixtures predominantly exhibit Raman features (the α and β modes) coming from the selected polyynes (*i.e.*, $HC_8H$ and $HC_{10}H$), as shown in Figure S.1 in the SI.

To collect UVRR spectra, we focused the synchrotron-based UV excitation approximately 7 mm above the graphite target surface. This geometry allowed us to avoid any interference with UVRR measurement coming from the plasma plume since the length of the plasma plume in our conditions is well below 3 mm, as observed in ablations at much higher fluences (tens of $J/cm^2$) [69,75–78] compared to our experiments. Moreover, in this way, we prevented any delay time issue since the travel time from the growth region to the focal point is of the order of a few *μs* since polyynes are ejected from the nucleation site at supersonic speed (approximately 1500 m/s) [73,75,78,79], carried by shockwaves generated after the collapse of the plasma plume.

To select suitable solvents for ablation experiments, we considered several factors critical to our study. Firstly, the UV transparency of the solvents within the beamline wavelength range of interest was vital, as it directly impacts the efficiency and accuracy of UVRR spectroscopy. Using UV-transparent solvents ensures the required penetration in the liquid environment of the UV Raman excitation and prevents any solvent-related absorption effects. Additionally, we aimed to explore how the solvents' polarity and C/H ratio influence polyynes' formation, dynamics, and stability during their synthesis by PLAL. Indeed, previous studies showed that solvents with reduced polarity showed enhanced polyynes' production with PLAL [37,38,80,81], while solvents with high C/H ratio promote the polymerization and the formation of longer chains [36,37,45,81,82]. For these reasons, we chose water, methanol, isopropanol, and acetonitrile. Water, with a UV cutoff at 190 nm, serves as a highly polar environment (relative polarity 1, as listed in Refs. [83,84]) and acts as a negative control to emphasize the role of organic solvents as additional carbon sources for the polymerization reaction of polyynes. Methanol, with a UV cutoff at 205 nm, is a slightly less polar solvent than water, with a relative polarity of 0.762 [83,84] and a C/H ratio of 0.33. Isopropanol, with a UV cutoff at 205 nm, has a lower polarity (0.564 [83,84]) and a higher C/H ratio (0.5). Acetonitrile, with a UV cutoff at 190 nm, presents the lowest polarity among the selected solvents (0.460 [83,84]) and the highest C/H ratio

(0.67). These solvents cover a broad spectrum of chemical properties, allowing us to systematically study solvent effects on polyynes' growth dynamics.

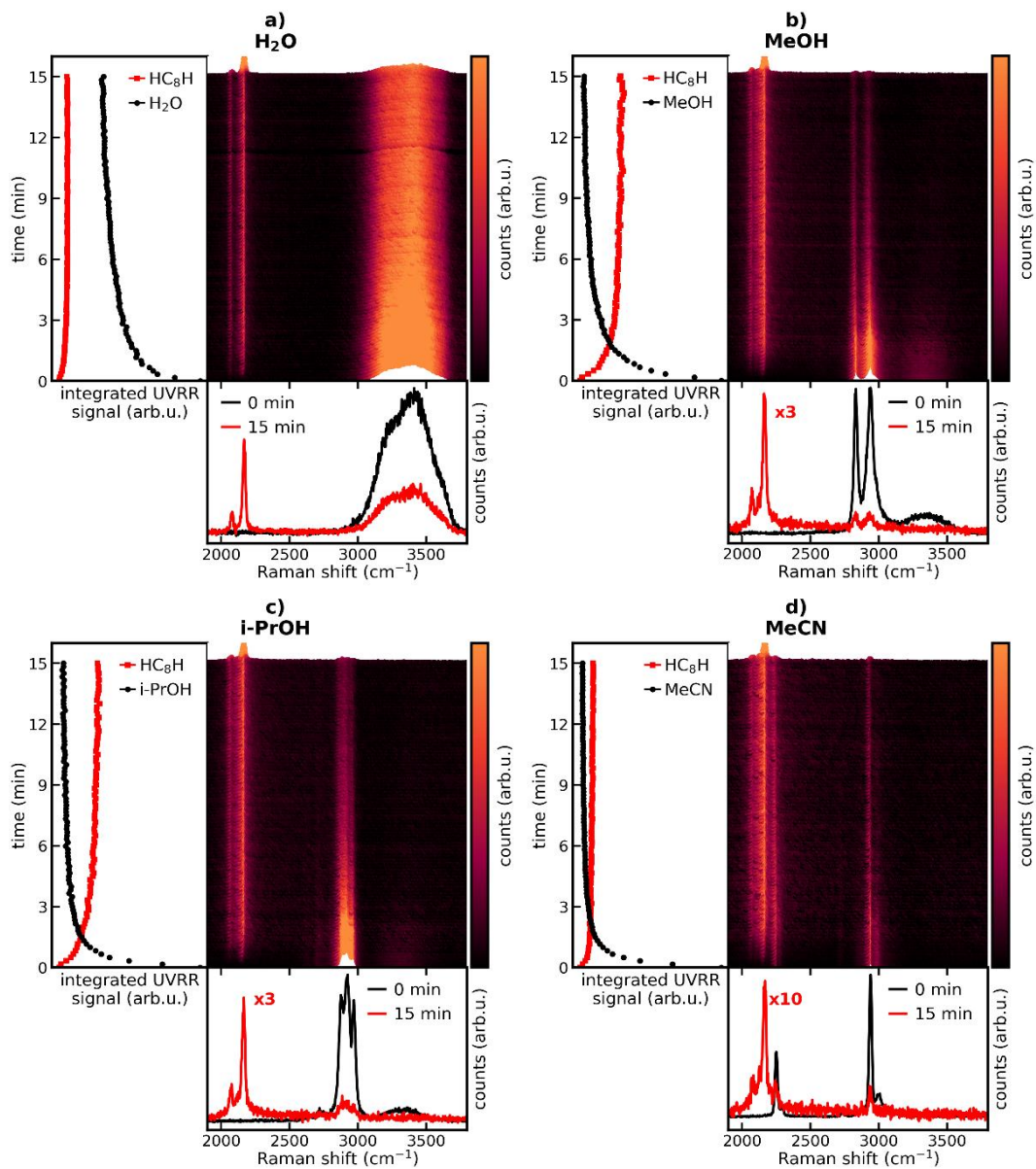

**Figure 2 In situ UVRR spectra at 226 nm during ablations in different solvents.** UVRR spectra collected at 226 nm Raman excitation in a) water, b) methanol (MeOH), c) isopropanol (i-ProH), and d) acetonitrile (MeCN) during 15 minutes of ablation (1064 nm ablation laser, 15 minutes of ablation time, 50 mJ per pulse), in the colormap of each panel. The integrated UVRR Raman signals of $HC_8H$'s α mode (red squares) and the relevant solvent Raman band (black circles, see main text) are displayed in the left-hand box of each panel. Fit errors (see Section S.1 in the SI) are shown with error bars. The first (0 min) and last (15 min) spectra are reported in the bottom box of each panel. The 15 min spectra are multiplied by a variable factor to improve the visualization.

Figure 2 shows the evolution of the *in situ* UVRR signals collected at 226 nm Raman excitation during ablation experiments in different solvents (water in panel a, methanol in panel b, isopropanol in panel c, and acetonitrile in panel d). We obtained analogous *in situ* UVRR dynamics at 251, 272, and 264 nm, shown in Figure S.2, S.3, and S.4 in the SI, respectively. We chose these wavelengths since they are in resonance with the most intense peak (0-0 vibronic absorption) of hydrogen-capped polyynes with 8, 10, and 12 sp-carbon atoms, *i.e.*, $HC_8H$ (226 nm), $HC_{10}H$ (251 nm), and $HC_{12}H$ (272 nm), as shown in Figure S.5 and Table S.1. We excited the hydrogen-capped polyyne with 14 sp-carbon atoms ($HC_{14}H$) at its 0-2 vibronic peak (*i.e.*, 264 nm) since this is the first available transition for this wire (*i.e.*, 0-0 at 295 nm and 0-1 at 280 nm, see Figure S.5) in the UV excitation range of the beamline (200-272 nm, see Method section). The ablation parameters selected in all experiments are listed in the Methods section.

UVRR spectra in Figure 2 (Raman excitation at 226 nm) display two main groups of signals at distinct frequency ranges. Below 2200 cm$^{-1}$, we observe two main bands at about 2175 and 2085 cm$^{-1}$ (averaging across different solvents; see Figure S.5b and Table S.2 in the SI), growing as ablation proceeds. Since we are in resonance with the 0-0 vibronic peak of HC$_8$H (Figure S.5a), these signals correspond to its α and β modes (Figure S.5b and Table S.2), in agreement with data in Refs. [25,70]. Moreover, the band at 2170 cm$^{-1}$ displays a less intense shoulder at lower frequencies (≈2120 cm$^{-1}$, averaging across different solvents, as displayed in Figure S.5b in the SI), visible for ablations in MeOH, i-PrOH, and MeCN (see Figure 2b, c, and d, respectively) at longer ablation times. We assigned this signal to the α mode of HC$_{10}$H (Table S.2), enhanced by matching its 0–2 vibronic transition at approximately 226 nm (Figure S.5a) [25,37,70,72,85]. A summary of the results presented herein is available in Table S.2 in the SI.

Correspondingly, *in situ* UVRR spectra collected at 251 nm (see Figure S.2 in the SI) and 272 nm (see Fig. S.3 in the SI) present features similar to what we observed at 226 nm. Notably, the α modes of HC$_{10}$H (2127 cm$^{-1}$, averaging across different solvents; see Figure S.5c and d in the SI) and HC$_{12}$H (2100 cm$^{-1}$, averaging across different solvents; see Figure S.5e and f in the SI) emerge as the predominant polyynic Raman peaks in the UVRR spectra recorded at 251 nm and 272 nm, respectively, according to literature data [25,70,72,85]. A relatively weak HC$_{12}$H signal was observed in water ablations, indicating reduced productivity of longer polyynes in this medium. Furthermore, we observe the presence of their β modes and weaker signals corresponding to HC$_{12}$H and HC$_{14}$H's α modes in 251 and 272 nm spectra, respectively. Refer to Table S.2 in the SI for a comprehensive overview of these results.

*In situ* measurements carried out at 264 nm reveal two distinct peaks at 2095 and 2060 cm$^{-1}$, corresponding to HC$_{12}$H and HC$_{14}$H's α modes (averaging across different solvents; see Figure S.5g and h in the SI), consistent with previous studies [25,37,70,72,85]. Detailed information is provided in Table S.2 in the SI. *In situ* UVRR spectra are provided for MeOH and MeCN, while no data is available for water due to the expected lower productivity for HC$_{14}$H compared to HC$_{12}$H, for which we recorded only a weak signal [34,37,51]. Moreover, spectra in i-PrOH are not reported due to the highly noisy background obscuring Raman signals, possibly originating from some plasma plume emissions particularly strong in ablations in i-PrOH, while they are weaker in ablations in MeOH and MeCN (see spectral disturbances above 3000 cm$^{-1}$ in Figure S.4 in the SI).

From these observations, this innovative approach demonstrates its unique ability to provide direct access to spectroscopic information related to single-chain species growth despite the presence in the ablated solutions of many different polyynes with different chain lengths and terminations and carbon-based byproducts like hydrocarbons [34,37,51,65]. This selectivity is unfeasible with other *in situ* techniques, like the surface-enhanced Raman scattering probe we developed in a previous work [73].

Regarding the other groups of signals present in UVRR spectra above 2800 cm$^{-1}$, they are assigned to the CH or OH stretching modes of the different solvents. Water displays its OH stretching band from approximately 2900 to 3800 cm$^{-1}$ (see Figure 2a). The same Raman band is visible, even if very weak, in alcohols (see Figure 2b and c) – *i.e.*, methanol and isopropanol. Instead, alcohols (Figure 2b and c) and acetonitrile (Figure 2d) exhibit intense Raman-active CH stretching modes between 2800 and 3000 cm$^{-1}$. Moreover, acetonitrile possesses a CN stretching vibration at about 2258 cm$^{-1}$.

To correctly evaluate the evolution of the growth of the main species selected by matching their resonant condition, we fit each spectrum with a multi-curve model (including solvent's signals) with a linear baseline correction. Regarding the solvents, we used different Raman bands to track their behaviors: we integrated the OH stretching band for water and the CH stretching modes for methanol, isopropanol, and acetonitrile. The fitting model is described in Section S.1 in the SI. The corresponding integrated UVRR signals (*i.e.*, the ascribed area of the Lorentzian curves) are displayed in the left box of each panel of Figure 2 by exciting at 226 nm – in resonance with the 0–0 vibronic peak of HC$_8$H – and Figure S.2, S.3, and S.4 in the SI by exciting at 251, 272, and 264 nm, respectively – in resonance with the 0–0 vibronic peak of HC$_{10}$H and HC$_{12}$H and the 0–2 vibronic peak of HC$_{14}$H, respectively.

The evolution of HC$_8$H's α mode and the solvent's Raman bands (see Figure 2) reveals a simultaneous increase in the intensity of polyynes peaks and a decrease in the Raman response of solvent peaks, consistently observed across all solvents. The decrease in solvent Raman intensity is particularly remarkable given that the solvent concentration did not vary during the ablations, and the focus condition remained unperturbed. This counterintuitive trend can be attributed to self-absorption (SA), a well-known phenomenon in resonance Raman experiments [86–88]. Typically, the SA effect reduces a sample's Raman response due to increased absorption at the Raman excitation wavelength. In our study, SA is not attributable to the solvents, as we deliberately selected UV-transparent solvents within the relevant beamline wavelength range. Instead, it arises from the concurrent production of polyynes and byproducts (like hydrocarbons) [36–40,45,47,48,57,65,73,89], which effectively absorb synchrotron-based Raman excitation. The SA is especially sensitive to polyyne's concentration, as we tuned the Raman excitation in resonance with a vibronic transition of each polyyne. Consequently, polyynes and byproducts absorb a portion of the UV power at the focal point within the mixture, enhancing SA and explaining the solvent's reduced Raman response.

Instead, the apparent saturation of polyynes' UVRR signal is caused by the raising mixture's SA and increased polyynes' Raman signal, both stemming from polyynes' concentration growth. However, *ex situ* data collected under similar ablation conditions in other studies demonstrated continuous growth of all the H-capped polyynes within the first 60 min of ablation in acetonitrile [73] and between 15 min (for $HC_{10}H$) and 25 min (for $HC_8H$) in water [66]. Therefore, we anticipate that *in situ* UVRR data will exhibit similar behavior, and a correction for SA is necessary to accurately evaluate polyynes' actual growth dynamics.

Addressing SA must be considered carefully because both the incoming synchrotron beam and the backscattered Raman photons are absorbed by the mixture, as shown by the UV-Vis absorption spectra of mixtures in Figure S.8 in the SI. We can distinguish slightly different absorbances experienced by the backscattered Raman photons of the α mode (from 2000 to 2200 $cm^{-1}$) or solvent Raman bands (CH stretching from 2800 to 3000 $cm^{-1}$ and OH stretching from 2900 to 3800 $cm^{-1}$). This contribution, however, is always weaker compared to the SA at the Raman excitation wavelength. Given the exponential nature of absorption, as a first approximation, we will consider the latter component as predominantly affecting *in situ* UVRR measurements.

We developed an empirical model to correct in situ UVRR data for SA, using the solvent Raman band as internal reference, as detailed in Section S.2 in the SI. As discussed earlier, the area of the Raman band of each solvent ($A_s(t)$, where $t$ is the ablation time) should remain equal to its initial value ($A_s(0)$). By multiplying this factor with the area of polyyne's α mode ($A_p(t)$), we obtain the SA-corrected integrated polyynes' signal ($A'_p(t)$):

$$A'_p(t) = A_p(t) \frac{A_s(0)}{A_s(t)} \qquad (Eq.\ 1)$$

The use of this empirical approach is justified by the unfeasibility to meet the requirements of analytical expression, as the model reported in Ref. [86], in terms of knowledge of physical aspects of polyynes and byproducts and geometrical details of the setup (see Section S.2 in the SI).

We can further analyze the corrected UVRR integrated polyynes' signal to obtain more significant information about the growth. Specifically, using *ex situ* UVRR measurements of size-selected H-capped polyynes with known concentration, we established a linear relationship between the integrated signal and concentration (see Section S.3 in the SI). This model links the integrated UVRR signal ($A'_{p,\lambda}(t)$) of each specific chain ($p$) collected at the characteristic excitation wavelength ($\lambda$) and ablation time ($t$) to its corresponding concentration ($c_{p,\lambda}(t)$) through this equation:

$$c_{p,\lambda}(t) = A'_{p,\lambda}(t) \cdot k_{\text{UVRR}\rightarrow c}(p, \lambda) \qquad (Eq.\ 2)$$

where we determined a conversion factor, $k_{UVRR \rightarrow c}(p, \lambda)$, from the calibration curve of each polyyne at each specific wavelength.

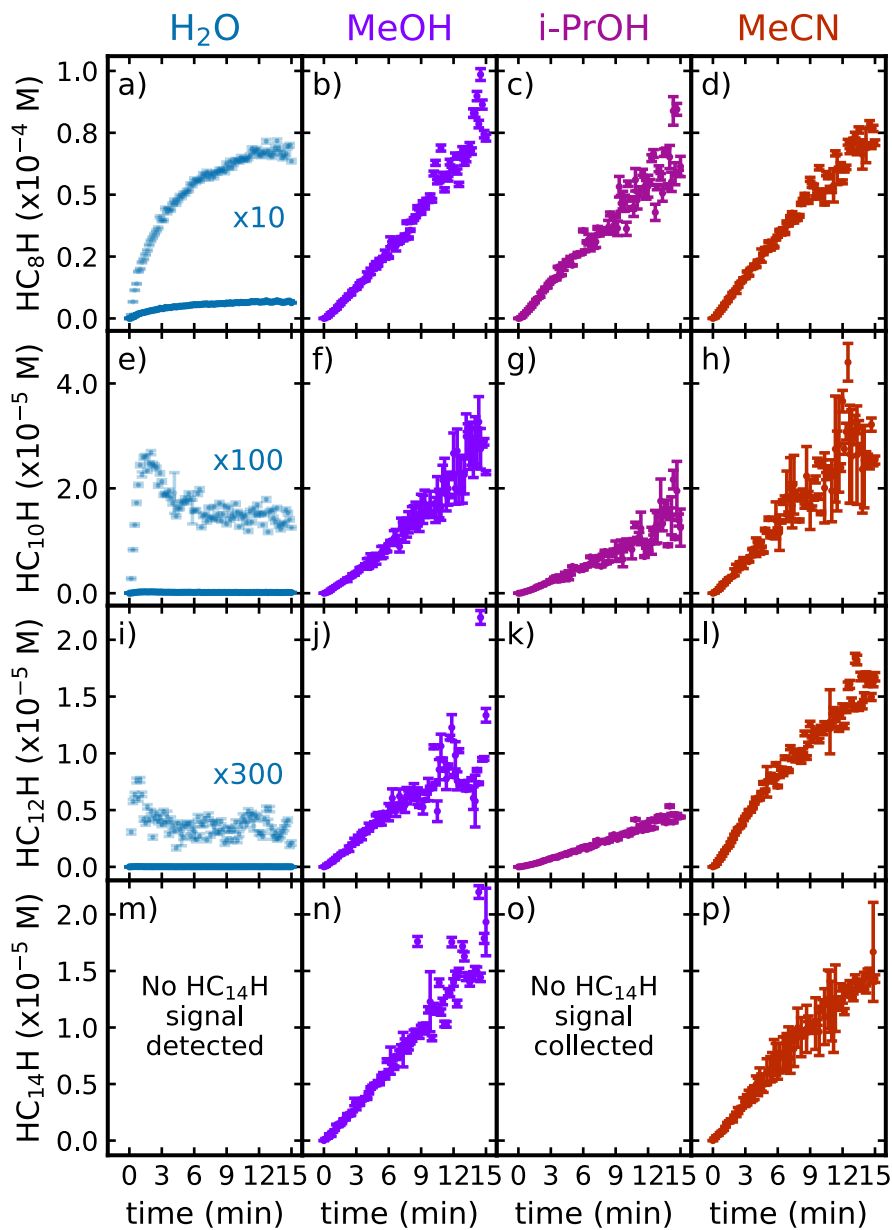

**Figure 3 Concentration evolution of H-capped polyynes during PLAL in various solvents.** Evolution of the concentration of size-selected H-capped polyynes during PLAL in different solvents as a function of the ablation time. The error bars derive from the fitting errors in determining α mode and Raman solvent's area and calculation of the conversion factor. Concentration behavior of $HC_8H$ extracted from SA-corrected *in situ* UVRR data collected at 226 nm in a) water, b) methanol, c) isopropanol, and d) acetonitrile. Concentration behavior of $HC_{10}H$ extracted from SA-corrected *in situ* UVRR data collected at 251 nm in e) water, f) methanol, g) isopropanol, and h) acetonitrile. Concentration behavior of $HC_{12}H$ extracted from SA-corrected *in situ* UVRR data collected at 272 nm in i) water, j) methanol, k) isopropanol, and l) acetonitrile. Concentration behavior of $HC_{14}H$ extracted from SA-corrected *in situ* UVRR data collected at 264 nm in n) methanol and p) acetonitrile. No signal of $HC_{14}H$ was detected in water (panel m), while we did not collect any signal at 264 nm in isopropanol (panel o). The concentrations of $HC_8H$, $HC_{10}H$, and $HC_{12}H$ during ablation in water were multiplied by, respectively, 10 (panel a), 100 (panel e), and 300 (panel i) to help their visualization.

Figure 3 illustrates the evolution of the concentration of size-selected H-capped polyynes during ablation in various solvents, *i.e.*, water, methanol, isopropanol, and acetonitrile. These results derive from *in situ* integrated UVRR signals, shown in Figure 2 and Figure S.2, S.3, and S.4 in the SI, as corrected for the SA effect using Eq. 1. The conversion into concentration was performed using Eq. 2.

We use concentration dynamics from ablations in water as a negative control to highlight the advantages of organic solvents and study possible degradation pathways. Indeed, our data confirms water as the least favorable environment for the growth of polyynes by PLAL, consistent with other *ex situ* data [36,37,90–92]. The concentrations reached in water for $HC_8H$, $HC_{10}H$, and $HC_{12}H$ were 1 or 2 orders of magnitude lower than in organic solvents (see Figure 3). This significant disparity underscores the importance of organic solvents as additional carbon sources for polyynes' formation. Specifically, water's inability to provide additional carbon atoms during plasma phase dissociation inhibits the polymerization reaction of polyynes [36,37,93]. A recent work studied cyanopolyyne's formation from an ablation in acetonitrile using $^{13}$C-enriched carbon powder and NMR analyses showed that more than one-fourth of the carbon in cyanopolyynes derives from solvent's dissociation [41]. Our results qualitatively confirm this finding, as water's production rate underperforms organic solvents, as described in the following.

Moreover, the high thermal conductivity of water (0.6062 W/m·K at 298 K) [94] compared to organic solvents such as MeOH (0.202 W/m·K), i-PrOH (0.135 W/m·K), and MeCN (0.188 W/m·K) [94] may contribute to the lower yield of polyynes observed in water ablations. This difference could result in a less confined plasma plume during the laser ablation, potentially reducing the formation rate of polyynes. Indeed, polyynes' synthesis requires strong out-of-equilibrium conditions and a high density of reactive carbon species that may be less favorable in a less dense plasma environment. Similar trends were observed with longer polyynes synthesized in propane gas than liquid hexane [81]. The lower yield in a gaseous environment can be attributed to the less dense plasma plume created by the lower pressure of propane gas compared to liquid hexane, as well as the lower thermal conductivity (propane gas 0.0185 W/m·K and liquid hexane 0.1167 W/m·K) [94]. However, the direct impact of thermal conductivity on polyynes' synthesis dynamics remains unclear [50], and further studies are needed to validate these hypotheses.

Interestingly, the growth of $HC_8H$ displayed in Fig. 3a in water follows a step-like response, *i.e.*, increasing in time and reaching a saturation concentration after approximately 12 min of ablation. We rationalized this behavior by fitting with the following equation:

$$c(t) = c_\infty \left[1 - e^{-t/\tau}\right] \quad \text{(Eq. 3)}$$

where $c_\infty$ (in mol/L) is the saturation concentration for $t \to \infty$ and $\tau$ (in min) is the characteristic time of the growth of this chain.

Instead, $HC_{10}H$ and $HC_{12}H$ exhibit a rapid first growth within the first 1-2 min of ablation, followed by a slower decrease towards a constant value, as illustrated in Figs. 3e and i. Then, the general equation for the growth of these two chains in water can be approximated as

$$c(t) = c_\infty \left[1 - e^{-t/\tau}\right] \cdot c_1 \left[1 + e^{-(t-t_0)/\tau_1}\right] \quad \text{(Eq. 4)}$$

where now $c_1$ (nondimensional), $t_0$ (min), and $\tau_1$ (min) are fitting parameters such that $c_\infty \cdot c_1$ is the new saturation concentration for $t \to \infty$. Results of the fit are reported in Figure S.12 in the SI.

The concentration of the different chains converges to comparable values at the end of the ablations, regardless of the organic solvent employed. However, distinct growth dynamics are observed when comparing alcohols and acetonitrile. Concerning ablations in alcohols (MeOH and i-PrOH), the production of polyynes by PLAL follows an almost linear increase during the entire 15 min of ablation, as shown in Figure 3. We used a linear model to represent these dynamics, given by

$$c(t) = \frac{c_\infty}{\tau} t \qquad \text{(Eq. 5)}$$

The slightly lower yield shown in i-PrOH for $HC_{10}H$ and $HC_{12}H$ may be related to a bathochromic effect, *i.e.*, the solvent-induced shift of polyynes' vibronic sequence towards longer wavelengths. UV-Vis spectra in Figure S.8 in the SI show the 0-0 peaks of $HC_{10}H$ and $HC_{12}H$ in i-PrOH at 252 nm and 276 nm, respectively, compared to 251 nm and 274 nm, respectively, in MeOH and MeCN. This minor bathochromic shift reduces the resonance enhancement and affects the absolute concentration values obtained from our conversion model.

In contrast, the growth dynamics of polyynes in MeCN deviate from linearity at longer ablation times, especially for longer chains, though these deviations are less pronounced than in water. For this reason, we modeled the growth of polyynes in MeCN with a single-characteristic-time model (Eq. 3) to extract the relevant growth parameters.

Figure S.12 in the SI presents the results of the fits applied to SA-corrected concentration dynamics in all solvents. In summary, Eq. 3 has been used to model the behavior of HC8H in water and HCnH (n=8-14) in MeCN, Eq. 4 can catch the dynamics of HC10H and HC12H in water, while Eq. 5 represents the evolution of HCnH (n=8-14) in MeOH and i-PrOH.

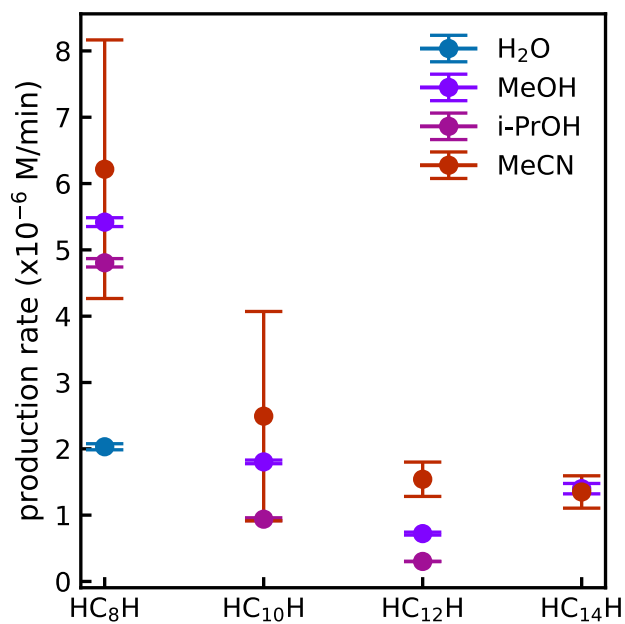

**Figure 4 Production rates of H-capped polyynes from PLAL in different solvents.** Production rates of H-capped polyynes (HC$_n$H, n=8-14) extracted from fitting procedures of the evolution of the corresponding concentration curves (see Figure S.12 in the SI). Numerical values are reported in Table S.4 in the SI. The error bars originate from the fit error. The production rates of HC$_{10}$H and HC$_{12}$H in water cannot be retrieved from the fitting model used.

The term $c_\infty/\tau$ in Eq. 5 can be interpreted as the production rate of H-capped polyynes, expressed in M/min (where M is molarity, *i.e.*, mol/L), which calculated values are displayed in Figure 4 (numerical values are reported in Table S.4 in the SI). This parameter provides an effective metric for evaluating the growth dynamics across various solvents. In the case of acetonitrile and water, when the growth is modeled using Eq. 3, we can approximate the single-characteristic-time model to its first-order Taylor expansion term, yielding the expression $c(t) \sim \frac{c_\infty}{\tau} t$ – *i.e.*, Eq. 5. However, this approximation does not hold for the growth dynamics of HC$_{10}$H and HC$_{12}$H in water. The production rate estimated for HC$_{14}$H in MeOH is larger than for HC$_{12}$H. This effect may be related to the more scattered concentration values for HC$_{14}$H, as reported in Figure 3n and Figure S.12m in the SI. The production rates in organic solvents are comparable, while those in water drop, confirming the low productivity when ablating in this environment. Excluding from our considerations the production rates of i-PrOH, for which the bathochromic effect introduces some distortions, and the scattered values for HC$_{14}$H, we notice a slight increase in the production rates passing from MeOH to MeCN.

The findings presented in Figures 3 and 4 enable us to review the growth dynamics of polyynes produced by PLAL across different solvents. The observed saturation at longer ablation times suggests that a peculiar formation-degradation equilibrium exists between the formation of polyynes by the PLAL and their degradation mechanisms. Degradation reactions are necessary to achieve this equilibrium since no evidence supports different production efficiencies at longer ablation times, as discussed in our previous works [36,73]. The most probable degradation pathways are oxidation and crosslinking reactions. We can exclude photo-induced degradation due to the lower energy of PLAL photons compared to polyynes' electronic transitions, while thermal-induced degradation is unlikely given the stable solution temperature of 300 K observed during similar experiments [36,37,73].

Instead, water's and alcohol's dissociation under laser irradiation can produce molecular, atomic, or radical oxygen, detrimental to polyynes' stability by inducing oxidation reactions [36,37,80,92,93]. Conversely, MeCN cannot dissociate in such species. Furthermore, MeCN possesses lower dissolved oxygen levels than the other solvents – one and two orders of magnitude less than water and alcohols, respectively [95,96]. However, the fast diffusion of molecular oxygen in water and organic solvents [97] prevents its accumulation, thus making it improbable to be responsible for long-term degradation reactions. Atomic and radical oxygen are highly reactive and react during the plasma phase (at most 100 ns after the ablation event) before polyynes can be detected by the *in situ* probe. These points indicate that degradation via oxidation is less significant, which is supported by the observed linear growth in alcohols where radical oxygen can be produced and molecular oxygen readily dissolves.

In contrast, the crosslinking probability increases with higher concentrations of polyynes and byproducts. In water, a highly polar solvent (relative polarity 1, where the polarity is defined as in Refs. [83,84]), the limited solubility of polyynes accelerates their aggregation, hastening degradation through crosslinking. This results in reaching the saturation concentration more rapidly, as observed for $HC_8H$ (Fig. 3a), than in organic solvents. For $HC_{10}H$ (Fig. 3e) and $HC_{12}H$ (Fig. 3i), this process leads to a saturation regime within less than 3 min, with a saturation concentration lower than the peak concentration.

Instead, polyynes exhibit higher solubility in organic solvents, whose relative polarity increases from MeCN (0.46) to i-PrOH (0.564) to MeOH (0.762) [83,84]. Despite MeCN's lower polarity compared to alcohols, it reaches the formation-degradation equilibrium faster. This accelerated equilibrium is due to the larger production rate of byproducts in MeCN compared to alcohols (Figure 4). MeCN's tendency to carbonization upon dissociation [37,82] and its higher C/H ratio (0.67) – *i.e.*, the balance between carbon and hydrogen atoms in solvent molecules – relative to alcohols (0.33 for MeOH and 0.5 for i-PrOH) contribute to this greater production rate. The solvent's C/H ratio plays a pivotal role in determining polyynes' nucleation [37,81,93,98]. These characteristics promote an extensive formation of carbonaceous compounds, including polyynes and byproducts, thereby enhancing the production rate while increasing crosslinking probability. The higher production of byproducts is evident by the dark brown color of MeCN mixtures, in contrast to yellow and orange solutions for MeOH and i-PrOH, respectively [37]. This heightened degradation in MeCN contrasts with our previous observation that identified MeCN as the most stable environment among those tested – water, MeOH, i-PrOH, ethanol, and MeCN – for polyynes' stability [37]. However, those results are based on ablations in MeCN using a 532 nm laser, which produces fewer byproducts than ablations at 1064 nm, indicated by a lighter MeCN mixture color (i.e., dark brown for 1064 nm and dark orange for 532 nm). Therefore, differences in production rates of both byproducts and polyynes can justify this discrepancy, emphasizing the importance of considering the wavelength effect on polyynes' growth dynamics in future studies.

Alcohols, as mentioned above, yield cleaner solutions with lower levels of byproducts. This reduces the probability of crosslinking reactions, resulting in linear growth (Figure 3). Despite their higher polarity than MeCN, crosslinking reactions show an enhanced correlation to the concentration of carbonaceous products within the mixture than increased aggregation due to high polarity. Like other solvents, we anticipate that alcohols will eventually reach a formation-degradation equilibrium, albeit at longer ablation times. The higher hydrogen content or lower C/H ratio in alcohols can stimulate termination reactions that stabilize polyynes while increasing the formation of byproducts such as hydrocarbons. Therefore, while the substantial C/H ratio of alcohols is pivotal in determining their high production rate (Figure 4), it is not the sole solvent property influencing this parameter.

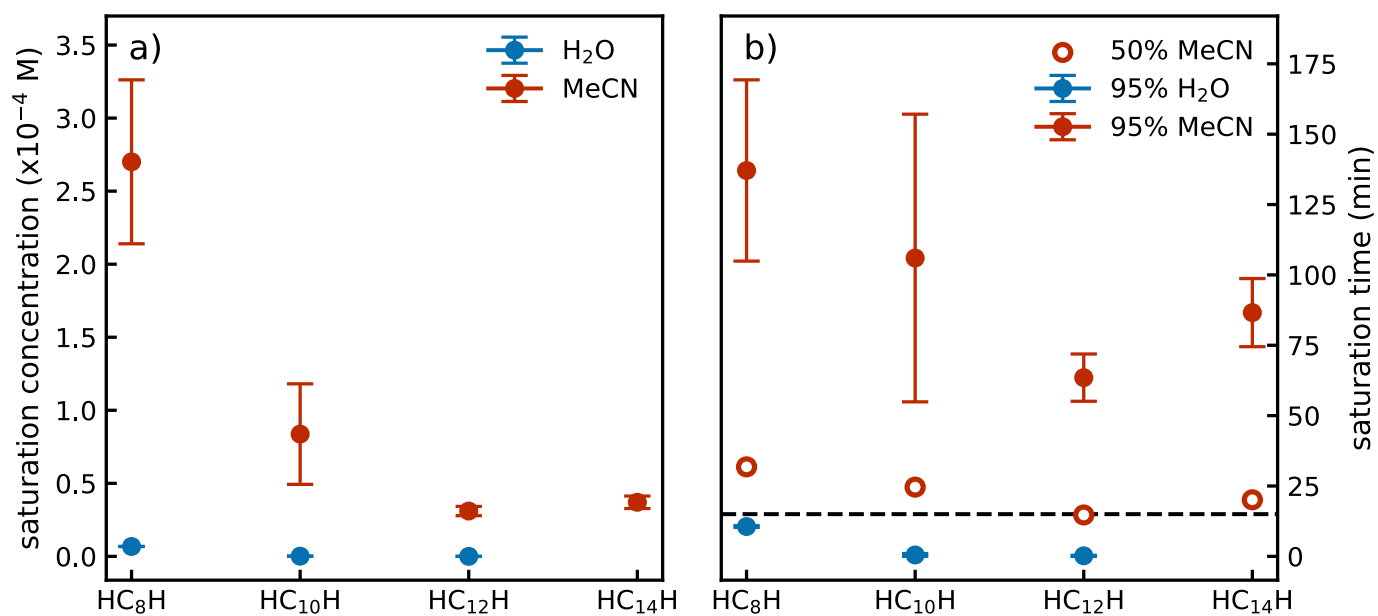

**Figure 5 Saturation concentration and time of H-capped polyynes from PLAL in water and acetonitrile.** a) Saturation concentration (in M) and b) saturation time to reach 50% (open circles) and 95% (filled circles) of the saturation concentration (in min) extracted from the fitting procedures of the evolution of the concentration curves in water and acetonitrile (see Figure S.12). The horizontal dashed line in panel b highlights 15 min value. Numerical values are reported in Table S.5 in the SI. The fit errors are shown as well.

For the solvents featuring formation-degradation equilibrium within 15 min of ablation (such as water) or showing a tendency towards it (like MeCN), we present in Figure 5a the saturation concentrations ($c_\infty$ from Eq. 3 or $c_\infty c_1$ from Eq. 4) achieved in our ablation conditions. Consequently, we display in Figure 5b two saturation times, defined as the ablation time needed to arrive at 50% (open circles) and 95% (filled circles) of the saturation concentration. These parameters serve as a practical measure to discuss polyynes' growth dynamics. The almost linear growth of shorter chains ($HC_8H$ and $HC_{10}H$) in MeCN introduces larger errors in determining the saturation concentration and time. The saturation concentration and saturation time for

HC$_{14}$H appear higher than HC$_{12}$H's, but these values might be inaccurately predicted due to the noisier *in situ* UVRR spectra recorded at 264 nm, as shown by the larger error bars in Figure 3p.

These parameters decrease with increasing chain length, reflecting the rapid reduction in the probability of longer polyynes' synthesis and their weaker stability as the chain length extends. They highlight even more the solvent role in providing additional carbon atoms. Indeed, the saturation concentrations achieved in ablations in acetonitrile overcome those in water by 2 or 3 orders of magnitude. In parallel, the saturation times to achieve 95% saturation concentration in water are relatively short, approximately 10 min for HC$_8$H and 26 and 8 s for HC$_{10}$H and HC$_{12}$H, respectively. These values mirror the insolubility and consequent instability of polyynes in water (see Fig. 5b and Table S.4 in the SI). Conversely, MeCN features much longer saturation times that easily overcome 1 h of ablation (see Figure 5b and Table S.4 in the SI). Instead, the ablation time to reach 50% of saturation concentration tells us that within the duration of our *in situ* UVRR experiments, we are close to these concentration values for all polyynes.

Concerning alcohols, *i.e.*, MeOH and i-PrOH, even though we cannot perform these kinds of estimations, we can predict their behavior at longer ablation times. Indeed, their lower tendency to carbonization suggests higher saturation concentrations and times in alcohols than those observed in MeCN.

Interestingly, we can predict the saturation concentration of longer H-capped polyynes, which cannot be probed due to the wavelength limitations of the beamline, starting from the data from ablations in MeCN reported in Figure 5a. Indeed, the size-dependent behavior of the saturation concentration in Fig. 5a can be accurately fit with an exponential function. Table S.6 in the SI reports the expected saturation concentrations for wires from HC$_{16}$H up to HC$_{30}$H, *i.e.*, the longest H-capped polyyne ever detected produced by PLAL [45], resulting in approximately 10$^{-9}$ mol/L. Despite the low concentration, the enhancement coming from resonance Raman and the increasing Raman activity of polyynes with the chain length (refer to the SI of Ref. [73]) could, in principle, allow us to detect this chain by collecting *in situ* UVRR spectra at 406 nm, *i.e.*, the wavelength of its 0-0 vibronic transition [45] However, this wavelength and the others needed to perform *in situ* UVRR experiments for chains longer than HC$_{14}$H are far beyond the available range of the IUVS beamline. Further experiments are required to assess the growth of longer chains as the new generation of tunable lasers may represent a valid alternative to provide sufficient tunability in this UV region (270 – 400 nm) approaching the visible range.

These results indicate the unnecessariness of longer ablations due to an upper limitation in polyynes' concentration in a limited volume mixture (*i.e.*, 2 mL in this work). This limit arises from degradation mechanisms, particularly crosslinking, which reduces the ablation efficiency over time. Therefore, alternative approaches should be explored to achieve polyynes' concentrations required for applications or characterizations with less sensitive techniques. These include performing ablations in larger liquid volumes or using continuous flow systems, thus applying post-ablation treatments, such as concentrating the solution to a smaller volume. Our findings also showed that solvents with a high C/H ratio and a tendency to carbonization upon dissociation tend to reach the formation-degradation equilibrium faster due to an enhanced production rate. This can result in poorer polyynes' storage stability and may necessitate filtration and purification steps after the ablation. Alternatively, transferring polyynes to stabilizing environments, such as encapsulating them in polymeric films or nanotubes, can stabilize them.

# Conclusions

The *in situ* UV resonance Raman approach presented in this study represents a unique and direct method to monitor carbon atomic wires' synthesis in liquid. It allows for real-time, direct monitoring of chains' growth during PLAL experiments without perturbing the synthesis process or the ablation environment. Exploiting the fine tunability of synchrotron-based UV radiation, we achieved unprecedented size selectivity, enabling us to track the single-chain species growth dynamics. Moreover, the sensitivity of the *in situ* UVRR probe, enhanced by matching the resonance condition, allowed us to detect and study the whole dynamics of wires' fabrication, even at the very early stages where their concentrations were down to $10^{-9}$-$10^{-10}$ mol/L.

We delved into the effect of diverse solvents in our *in situ* PLAL experiments, including water, methanol, isopropanol, and acetonitrile. This comprehensive and systematic approach allowed us to gain unprecedented insights into carbon atomic wires' growth dynamics through PLAL. By developing an empirical model to correct UVRR data from self-absorption and converting them into concentration values, we gained a deeper understanding of these dynamics.

Our modeling of the growth dynamics of carbon atomic wires during PLAL revealed size-dependent production rates, which range from ca. $10^{-6}$ to $10^{-7}$ M/min, meaning a size-selected H-capped polyyne average production of approximately $10^{11}$-$10^{12}$ wires per pulse with a 10 Hz laser (*i.e.*, 600 shots per minute) in 2 mL of solution. These production rates are influenced by the solvent's properties, with carbon-rich solvents that tend to carbonize under dissociation resulting in higher production rates.

We unveiled that the growth did not follow a linear behavior. Instead, the concentration of carbon atomic wires tends to reach an equilibrium state between formation and degradation, primarily due to crosslinking reactions. This equilibrium results in a saturation concentration that decreases and is reached more quickly as chain length increases, indicating a reduced synthesis yield and stability for longer chains.

Organic solvents yielded saturation concentrations that were orders of magnitude higher and took significantly longer times to reach equilibrium compared to water. We think that future studies should consider using $^{13}$C-enriched solvents as a quantitative way to evaluate the solvent's contribution in providing additional carbon atoms for polyynes polymerization. Our findings demonstrated that performing single, longer (> 1 hour) ablations in small liquid volumes is not effective for obtaining high concentration of carbon atomic wires through PLAL. Instead, alternative approaches must be explored to overcome these limitations.

Our findings demonstrate the importance of *in situ* characterization techniques in providing valuable insights into the growth dynamics of carbon atomic wires and, more in general, in enhancing our ability to study and manipulate materials at the atomic and molecular levels. In particular, the exceptional versatility and scalability of PLAL provides the foundation for tailoring sp-carbon chains' structure and achieving scale-up production of carbon atomic wires by PLAL. This will unlock the possibility of applying sp-carbon chains in a wide range of applications beyond the existing prototypal devices [99–103] that showcase the outstanding potential of this material.

# Experimental Section

## Pulsed laser ablation in liquid

We performed pulsed laser ablation in liquid (PLAL) using a ns-pulsed Nd:YAG laser (Quantel Q-Smart 850, repetition rate 10 Hz), as shown in Figure 1. We employed the fundamental harmonic of the Nd:YAG laser at 1064 nm, with a pulse duration of 5 ns. We fixed the laser energy to 50 mJ per pulse through a beam attenuator module equipped with the laser head. We focused the laser beam with a lens with a 200 mm focal length (see Figure 1b) onto a graphite target (Testbourne Ltd., purity 99.99%). We performed ablations for 15 minutes in 2 mL of different solvents: deionized water ($H_2O$) Milli-Q (conductivity 0.055 µS), methanol (MeOH, Sigma-Aldrich, purity ≥99.9%), isopropanol (i-PrOH, Sigma-Aldrich, purity ≥99.9%), and acetonitrile (MeCN, Sigma-Aldrich, purity ≥99.9%).

We placed the target and the solvent into a PTFE ablation cell, prepared *ad hoc* for these experiments and schematically reported in Figure 1b. The cell features a cylindrical hole, opened at the top, allowing the PLAL laser to enter, and two quartz windows (Hellma, QS UV-Vis range 200-2500 nm), through which we focused the monochromatized synchrotron radiation to collect UV resonance Raman spectra during ablation experiments. The liquid volume reached a height of ≈18.7 mm from the target surface inside the ablation cell, enough to completely cover the quartz windows. We set the target-to-lens distance to approximately 177.9 ± 0.6 mm by fixing the ablation cell on a steel pillar.

## UV resonance Raman spectroscopy

*In situ* UV resonance Raman (UVRR) spectra were acquired using the synchrotron-based UVRR setup accessible at the BL10.2-IUVS beamline within Elettra Sincrotrone (Trieste, Italy) and described in detail here [104]. Various excitation wavelengths in the deep UV range were employed, as detailed in Table S.1 of the Supporting Information (SI), by precisely adjusting the undulator gap aperture to tune the energy of the emitted synchrotron radiation (SR). Subsequently, the SR light was monochromatized *via* a 750 cm focal length spectrograph equipped with a holographic grating featuring 3600 grooves/mm. It is worth noting that UV excitation beyond 272 nm or below 200 nm is not possible due to the limitations imposed by the minimum undulator gap aperture and by the optical elements of the monochromator. The incident radiation power on the samples varied, ranging from a few to tens of µW (see Table S.1 in the SI). Spectrometer calibration was executed using the spectrum of cyclohexane (spectroscopic grade, Sigma Aldrich). The ultimate spectral resolution was influenced by several factors, including the SR monochromator's resolving power (determined by factors such as focal length, grating, and slit aperture), the excitation wavelength, and the spectral range as constrained by the pixel spacing of the CCD. For each excitation wavelength, the final resolution of the recorded Raman spectra could be reliably estimated using the following general equation $\frac{spectral\ range\ [cm^{-1}]}{1340}$. As an illustration, resolutions of 2.6, 1.9, and 1.6 cm$^{-1}$/pixel were achieved at 216, 251, and 272 nm excitation wavelengths, respectively. The spectra were all collected at room temperature.

## Acknowledgements

P.M., S.P., S.M., A.L.B., V.R., and C.S.C. acknowledge funding from the European Research Council (ERC) under the European Union's Horizon 2020 research and innovation program ERC Consolidator Grant (ERC CoG2016 EspLORE grant agreement no. 724610, website: www.esplore.polimi.it). We acknowledge Elettra Sincrotrone Trieste for providing access to its synchrotron radiation facilities and for financial support under the SUI internal project (proposal numbers 20215090). The authors acknowledge the CERIC-ERIC Consortium for the access to experimental facilities and financial support (proposal number 20227204).

## Author contributions

P.M., S.P., A.L.B., V.R., and C.S.C. conceived the experiment. P.M., S.P., B.R., A.G., and C.S.C carried out in situ UV resonance measurements. A.G. designed and realized the ablation cell. P.M., S.P., S.M., B.R., V.R., and carried out ex situ UV resonance Raman measurements and all the other ex situ measurements. P.M., S.P., and S.M. performed data analysis. P.M. wrote the first version of the paper. All authors discussed the results and contributed to writing subsequent manuscript drafts.

## Competing Interest

The authors declare no competing interests.

## Data Availability

The data that support the findings of this study are available from the corresponding author upon reasonable request. They have been deposited in Zenodo with this identifier https://doi.org/10.5281/zenodo.10843187.

# Supporting Information

# Exploring the Growth Dynamics of Size-selected Carbon Atomic Wires with *in situ* UV Resonance Raman Spectroscopy

Pietro Marabotti[a,b*], Sonia Peggiani[a], Simone Melesi[a], Barbara Rossi[c], Alessandro Gessini[c], Andrea Li Bassi[a], Valeria Russo[a], Carlo Spartaco Casari[a*]

[a] Department of Energy, Micro and Nanostructured Materials Laboratory - NanoLab, Energy, Politecnico di Milano, Via Ponzio 34/3, Milano 20133, Italy
[b] Institut für Physik, Humboldt Universität zu Berlin, 12489 Berlin, Germany
[c] Elettra Sincrotrone Trieste, S.S. 114 km 163.5, Basovizza, 34149 Trieste, Italy.

*Corresponding authors: pietro.marabotti@polimi.it, carlo.casari@polimi.it

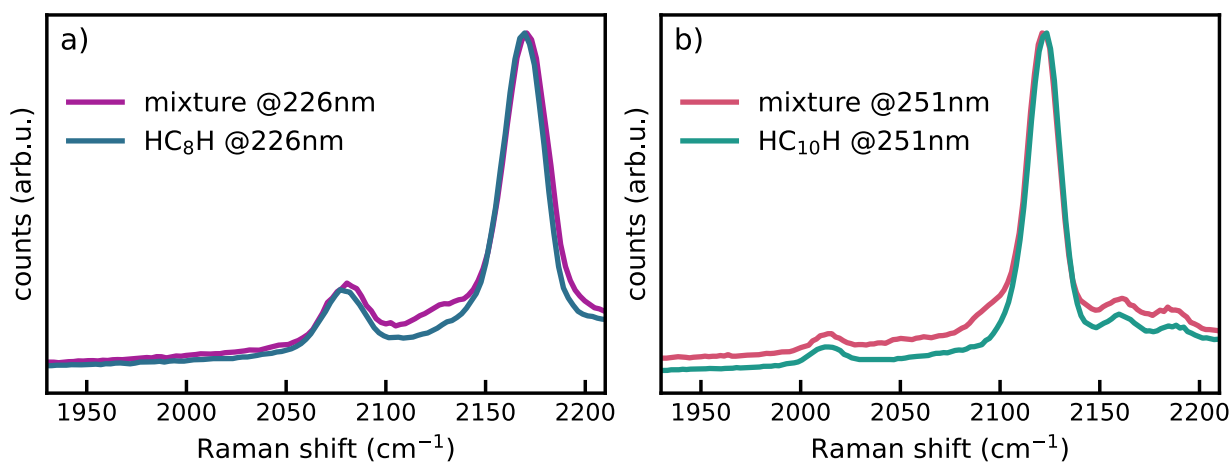

**Figure S1 Comparison between mixture and size-selected UVRR spectra of polyynes.** a) UVRR spectra of a mixture of polyynes (purple curve) and a size-selected H-capped polyyne ($HC_8H$, blueish curve) collected at 226 nm as excitation wavelength. b) UVRR spectra of a mixture of polyynes (pinkish curve) and a size-selected H-capped polyyne ($HC_{10}H$, blueish curve) collected at 251 nm as excitation wavelength.

| Polyyne | Excitation wavelength [nm] | Power on the sample [µW] | Aperture slits [µm] | Acquisition time [s] | CN stretching mode area (pristine solution) |
|---|---|---|---|---|---|
| $HC_8H$ | 226 | 16.9 | 50 | 10 | 27178 ± 67 |
| $HC_{10}H$ | 251 | 18.6 | 50 | 10 | 14094 ± 28 |
| $HC_{12}H$ | 272 | 8.7 | 30 | 10 | 34832 ± 51 |
| $HC_{14}H$ | 264 | 14.7 | 30 | 10 | 90861 ± 111 |

**Table S1** *In situ* UVRR parameters employed to monitor polyynes' growth during pulsed laser ablation in liquid (PLAL) experiments for each polyyne chain. CN stretching mode's UVRR integrated signal extracted before ablations in acetonitrile is reported.

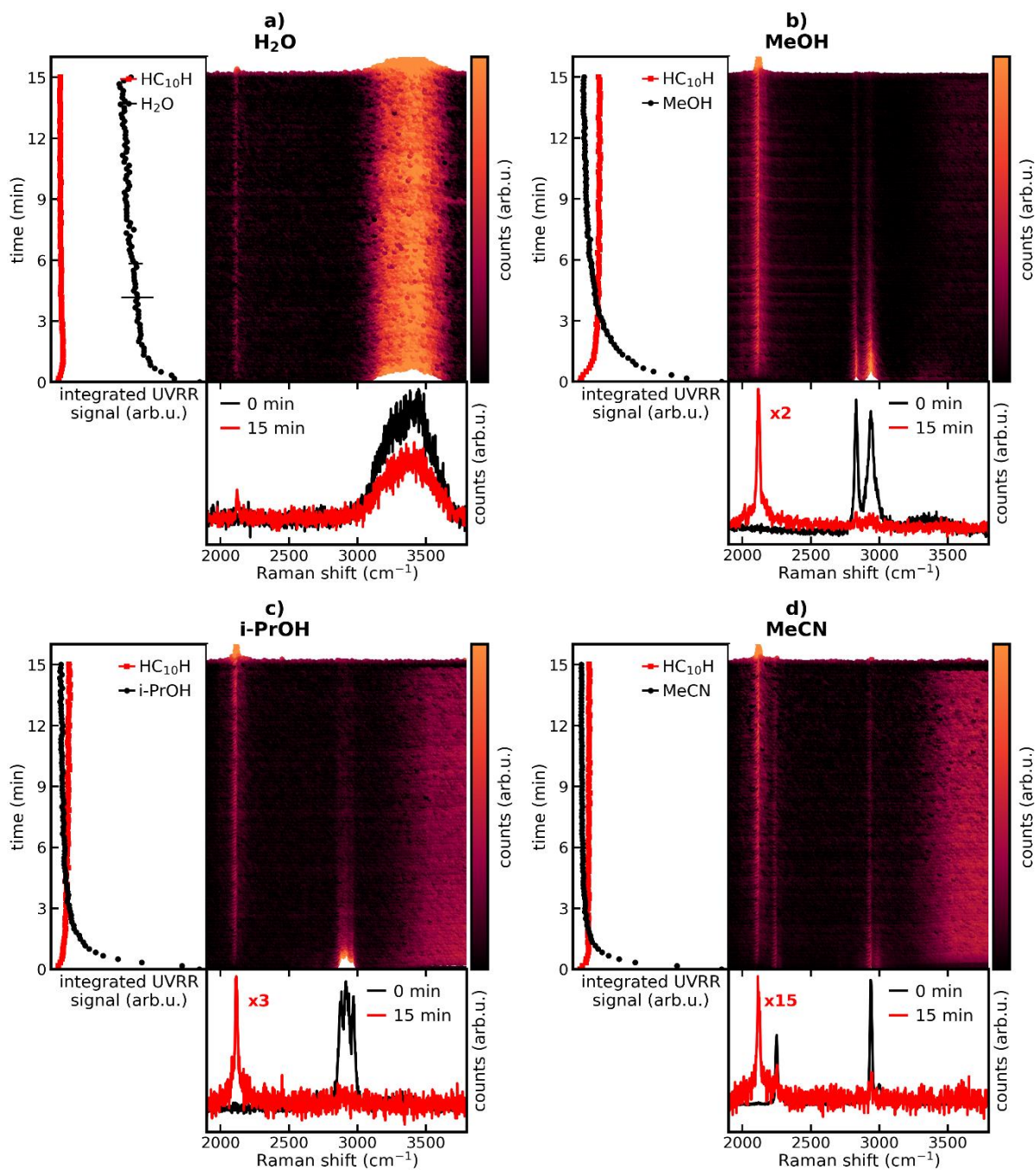

**Figure S2 In situ UVRR spectra at 251 nm during ablations in different solvents.** UVRR spectra collected at 251 nm Raman excitation in a) water, b) methanol (MeOH), c) isopropanol (i-ProH), and d) acetonitrile (MeCN) during 15 minutes of ablation (1064 nm ablation laser, 15 minutes of ablation time, 50 mJ per pulse), in the colormap of each panel. The integrated UVRR Raman signals of $HC_{10}H$'s α mode (red squares) and the relevant solvent Raman band (black circles, see main text) are displayed in the left-hand box of each panel. Fit errors (see Section S1) are shown with error bars. The first (0 min) and last (15 min) spectra are reported in the bottom box of each panel. The 15 min spectra are multiplied by a variable factor to improve the visualization.

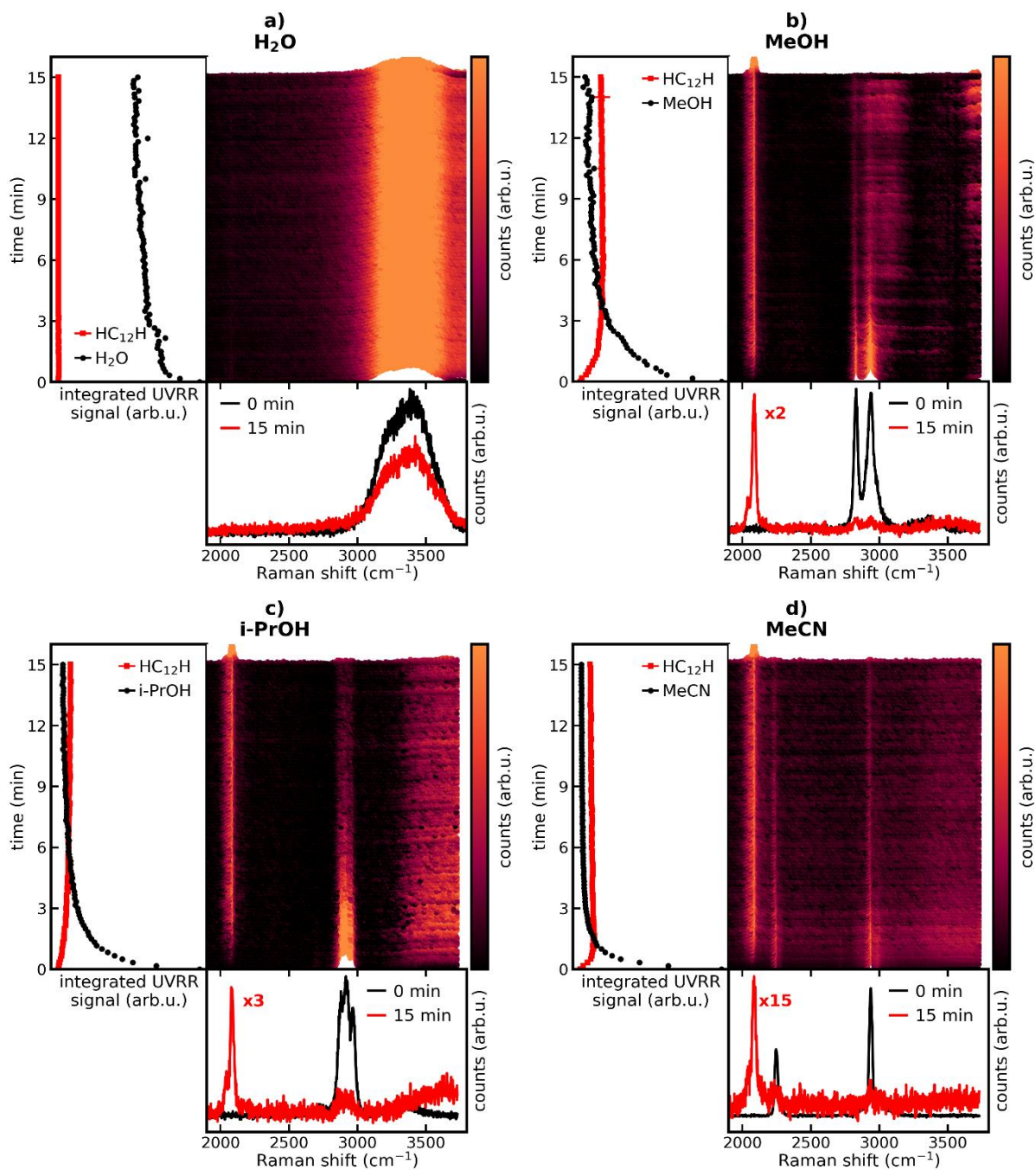

**Figure S3 In situ UVRR spectra at 272 nm during ablations in different solvents.** UVRR spectra collected at 272 nm Raman excitation in a) water, b) methanol (MeOH), c) isopropanol (i-ProH), and d) acetonitrile (MeCN) during 15 minutes of ablation (1064 nm ablation laser, 15 minutes of ablation time, 50 mJ per pulse), in the colormap of each panel. The integrated UVRR Raman signals of $HC_{12}H$'s α mode (red squares) and the relevant solvent Raman band (black circles, see main text) are displayed in the left-hand box of each panel. Fit errors (see Section S1) are shown with error bars. The first (0 min) and last (15 min) spectra are reported in the bottom box of each panel. The 15 min spectra are multiplied by a variable factor to improve the visualization.

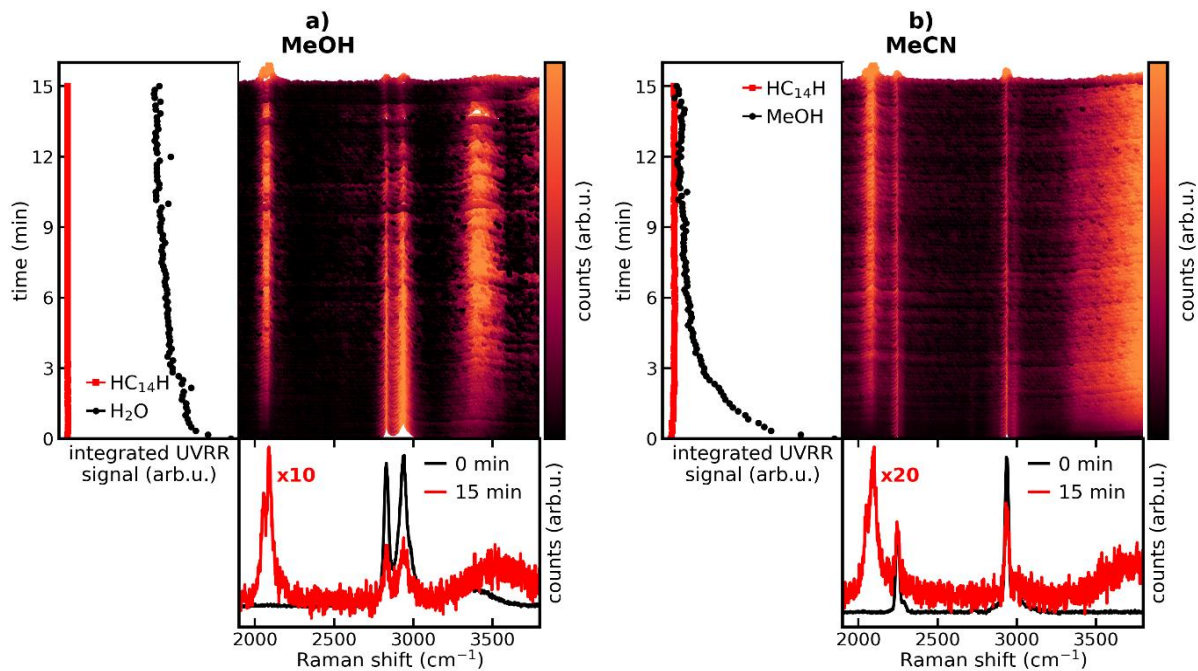

**Figure S4 In situ UVRR spectra at 264 nm during ablations in different solvents.** UVRR spectra collected at 264 nm Raman excitation in a) methanol (MeOH) and b) acetonitrile (MeCN) during 15 minutes of ablation (1064 nm ablation laser, 15 minutes of ablation time, 50 mJ per pulse), in the colormap of each panel. The integrated UVRR Raman signals of $HC_{14}H$'s α mode (red squares) and the relevant solvent Raman band (black circles, see main text) are displayed in the left-hand box of each panel. Fit errors (see Section S1) are shown with error bars. The first (0 min) and last (15 min) spectra are reported in the bottom box of each panel. The 15 min spectra are multiplied by a variable factor to improve the visualization.

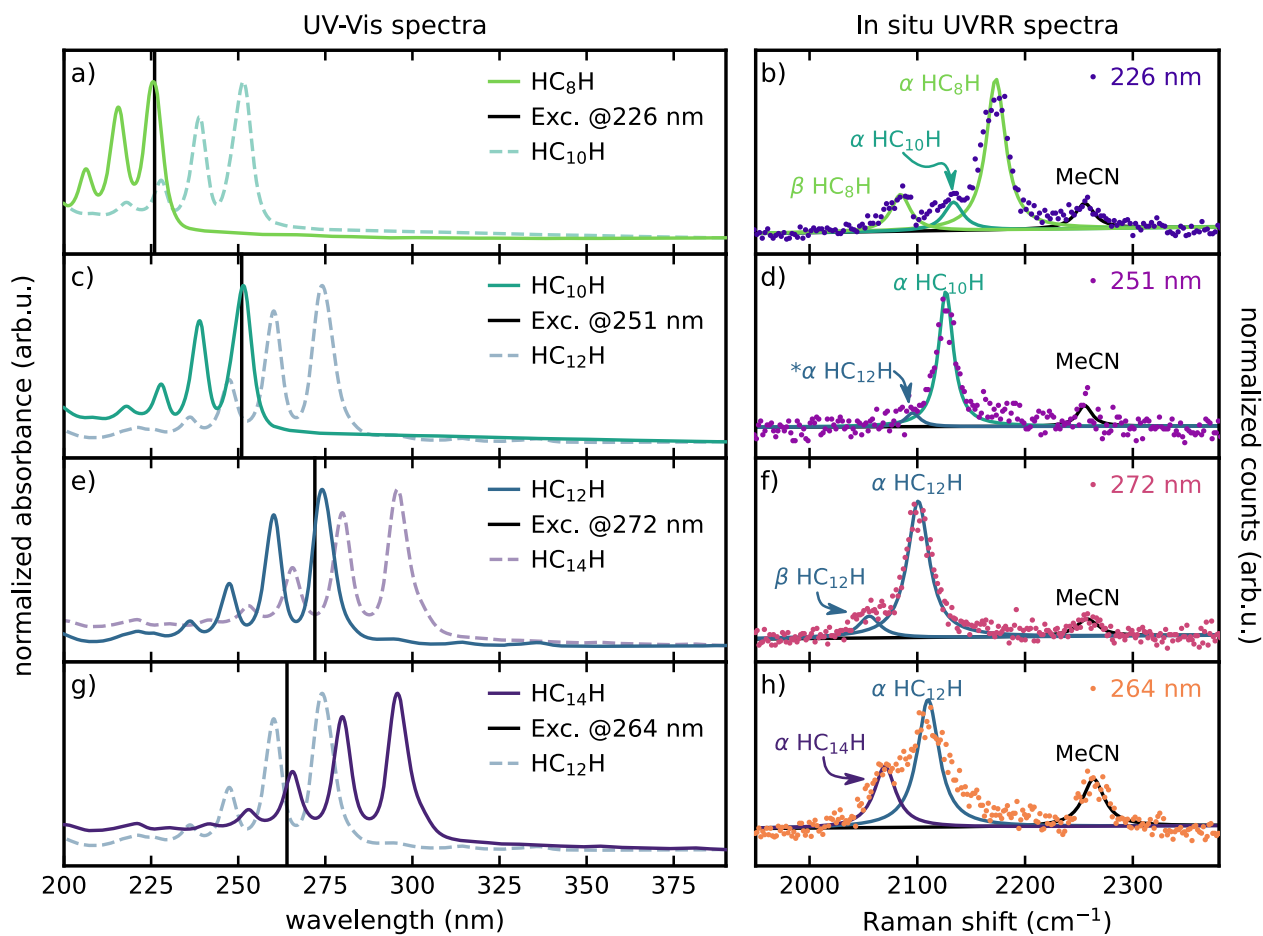

**Figure S5 Resonance conditions from UV-Vis spectra and assignment of spectra features in UVRR spectra of H-capped polyynes.** a) UV-Vis spectra of $HC_8H$ (solid line) and $HC_{10}H$ (semi-transparent dashed line). The black solid line at 226 nm, crossing the 0-0 and 0-2 vibronic peaks of $HC_8H$ and $HC_{10}H$, respectively, indicates the Raman excitation wavelength employed to record *in situ* UVRR spectrum in panel b. b) *In situ* UVRR spectrum collected at Raman excitation 226 nm during PLAL in acetonitrile (dark purple dots). The labels mark the Raman peaks present in the spectrum. The fitted Lorentzian Raman features are reported as solid colored lines. c) UV-Vis spectra of $HC_{10}H$ (solid line) and $HC_{12}H$ (semi-transparent line). The black solid line at 251 nm, crossing the 0-0 and approximately the 0-2 vibronic peaks of $HC_{10}H$ and $HC_{12}H$, respectively, indicates the Raman excitation wavelength employed to record *in situ* UVRR spectrum in panel d. d) *In situ* UVRR spectrum collected at Raman excitation 251 nm during PLAL in acetonitrile (purple dots). The labels mark the Raman peaks present in the spectrum. The α mode of $HC_{12}H$ is barely visible (marked with an "*"). The fitted Lorentzian Raman features are reported as solid colored lines. e) UV-Vis spectra of $HC_{12}H$ (solid line) and $HC_{14}H$ (semi-transparent dashed line). The black solid line at 272 nm, crossing the 0-0 and in between the 0-1 and 0-2 vibronic peaks of $HC_{12}H$ and $HC_{14}H$, respectively, indicates the Raman excitation wavelength employed to record *in situ* UVRR spectrum in panel f. f) *In situ* UVRR spectrum collected at Raman excitation 272 nm during PLAL in acetonitrile (dark pink dots). The labels mark the Raman peaks present in the spectrum. The fitted Lorentzian Raman features are reported as solid colored lines. g) UV-Vis spectra of $HC_{14}H$ (solid line) and $HC_{12}H$ (semi-transparent dashed line). The black solid line at 264 nm, crossing the 0-2 and approximately the 0-1 vibronic peaks of $HC_{14}H$ and $HC_{12}H$, respectively, indicates the Raman excitation wavelength employed to record *in situ* UVRR spectrum in panel h. h) *In situ* UVRR spectrum collected at Raman excitation 264 nm during PLAL in acetonitrile (orange dots). The labels mark the Raman peaks present in the spectrum. The fitted Lorentzian Raman features are reported as solid colored lines.

| Raman excitation [nm] | Solvent | Detected Raman shift [cm$^{-1}$] | | |
|---|---|---|---|---|
| | | β HC$_8$H (weak) | α HC$_{10}$H (shoulder) | α HC$_8$H (strong) |
| 226 | Water | 2089 | – | 2179 |
| | MeOH | 2083 | 2130 | 2175 |
| | i-PrOH | 2084 | 2132 | 2175 |
| | MeCN | 2086 | 2134 | 2176 |
| | | β HC$_{10}$H | α HC$_{12}$H | α HC$_{10}$H |
| 251 | Water | 2018 | – | 2130 |
| | MeOH | 2019 | 2101 | 2127 |
| | i-PrOH | 2018 | 2100 | 2126 |
| | MeCN | 2021 | 2099 | 2128 |
| | | β HC$_{12}$H | α HC$_{14}$H | α HC$_{12}$H |
| 272 | Water | – | – | 2106 |
| | MeOH | 2052 | – | 2101 |
| | i-PrOH | 2053 | – | 2099 |
| | MeCN | 2054 | – | 2100 |
| | | β HC$_{12}$H | α HC$_{14}$H | α HC$_{12}$H |
| 264 | MeOH | – | 2064 | 2102 |
| | MeCN | – | 2066 | 2106 |

**Table S2** Average Raman shifts of the Raman modes detected in the polyyne frequency range (1800–2200 cm$^{-1}$) in *in situ* UVRR data of Figure 2 (see main text), Figure S2, Figure S3, and Figure S4. The "–" symbol indicates that a mode is not observed in the corresponding UVRR spectra. The frequencies are rescaled by a common factor for each Raman excitation wavelength, *i.e.*, 1.00482 for 226 nm, 1.00397 for 251 nm, 1.00637 for 272 nm, and 1.00525 for 264 nm. This factor makes the CN stretching mode of MeCN match its tabulated frequency, *i.e.*, 2258 cm$^{-1}$.

## S1 Fitting procedure

We employed a custom fitting procedure to model *in situ* UVRR spectra and evaluate the evolution of the different Raman bands during the ablation.

We first identify the best fitting function for each pristine solvent (see Figure S6). For the OH stretching band of water, we employed two Gaussian curves, while we used only one Gaussian curve for the OH band in alcohols (methanol and isopropanol). We selected a Lorentzian function for each peak in the CH stretching frequency region of organic solvents (methanol, isopropanol, and acetonitrile). We chose the same function for the CN stretching mode of acetonitrile. The number of curves used to fit the solvent's Raman features has been selected to estimate their area as well as possible without completely modeling the corresponding Raman bands. To fit the solvent's Raman peaks during *in situ* UVRR experiments, we fixed the full width at half maximum (FWHM) of each solvent's Raman peak to the results obtained for the pristine solution. This prevents overfitting caused by numerous peaks in each spectrum.

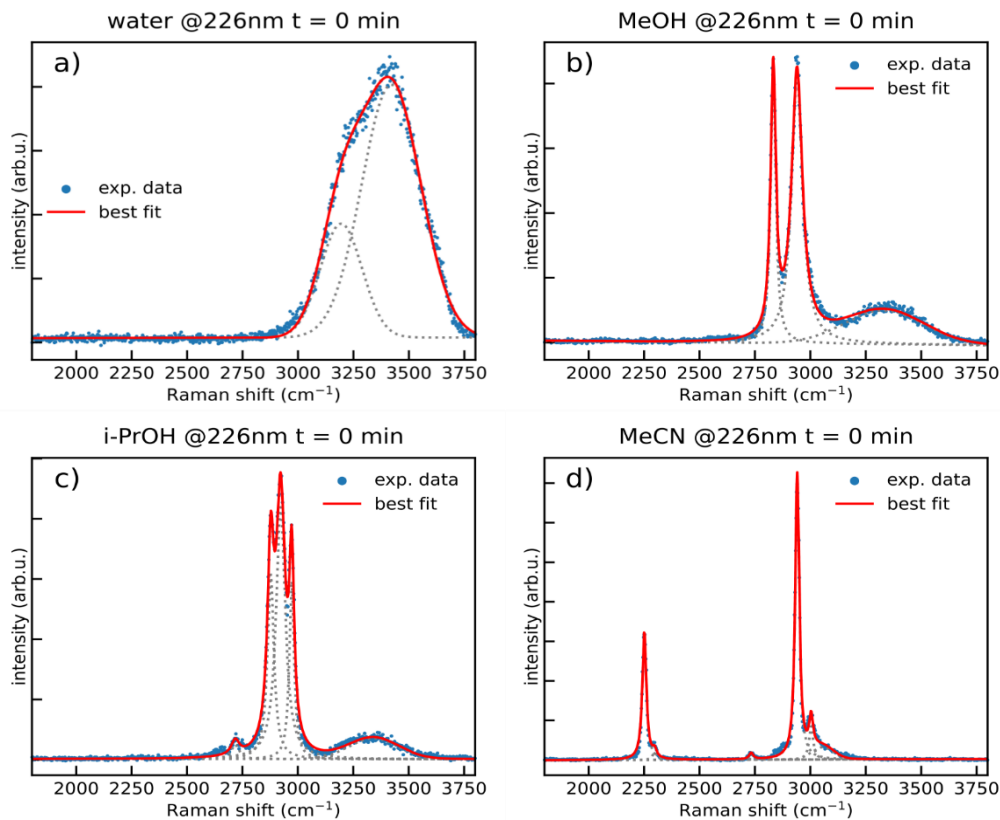

**Figure S6 UVRR spectra at 226 nm of pristine solvents.** UVRR spectra collected at 226 nm of pristine a) water, b) methanol, c) isopropanol, and d) acetonitrile solutions (blue dots). The red line represents the best fit function of each spectrum. Gray dotted lines show each component of the best fit.

In each series of *in situ* UVRR spectra, *i.e.*, excited at the same wavelength, we used a Lorentzian curve for all the polyynes' Raman modes, namely the α and β modes of each $HC_nH$ polyyne excited at its 0–0 vibronic transition ($HC_8H$ in Figure S5) and the α mode of the $HC_{n+2}H$ polyyne excited at its 0–2 vibronic transition ($HC_{10}H$ in Figure S5). In particular, for each excitation wavelength, we set the FWHM of polyynes' Raman peaks to that of the CH stretching vibration of acetonitrile at approximately 2940 $cm^{-1}$ and did not vary them during the fitting of *in situ* UVRR data. This allowed us to use the area as the parameter of merit to monitor the evolution of polyynes' α mode during the ablation. Figure S7 shows the results of the fitting process after 5 min of ablations in the case of *in situ* UVRR spectra collected at 226 nm, *i.e.*, in resonance with the 0-0 vibronic of $HC_8H$.

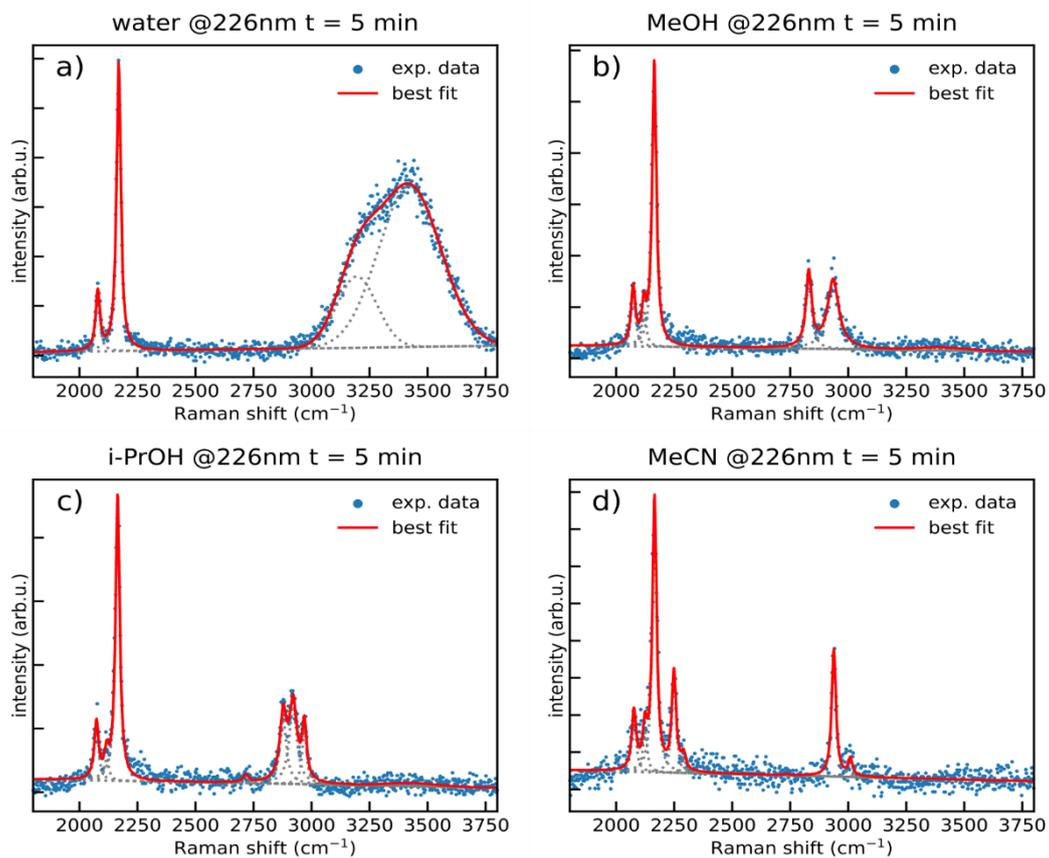

**Figure S7 In situ UVRR spectra at 226 nm after 5 min of ablations in different solvents.** *In situ* UVRR spectra collected at 226 nm after 5 min of ablation in a) water, b) methanol, c) isopropanol, and d) acetonitrile (blue dots). The red line represents the best fit function of each spectrum. Gray dotted lines show each component of the best fit.

# S2 Correction of self-absorption in resonance Raman spectra

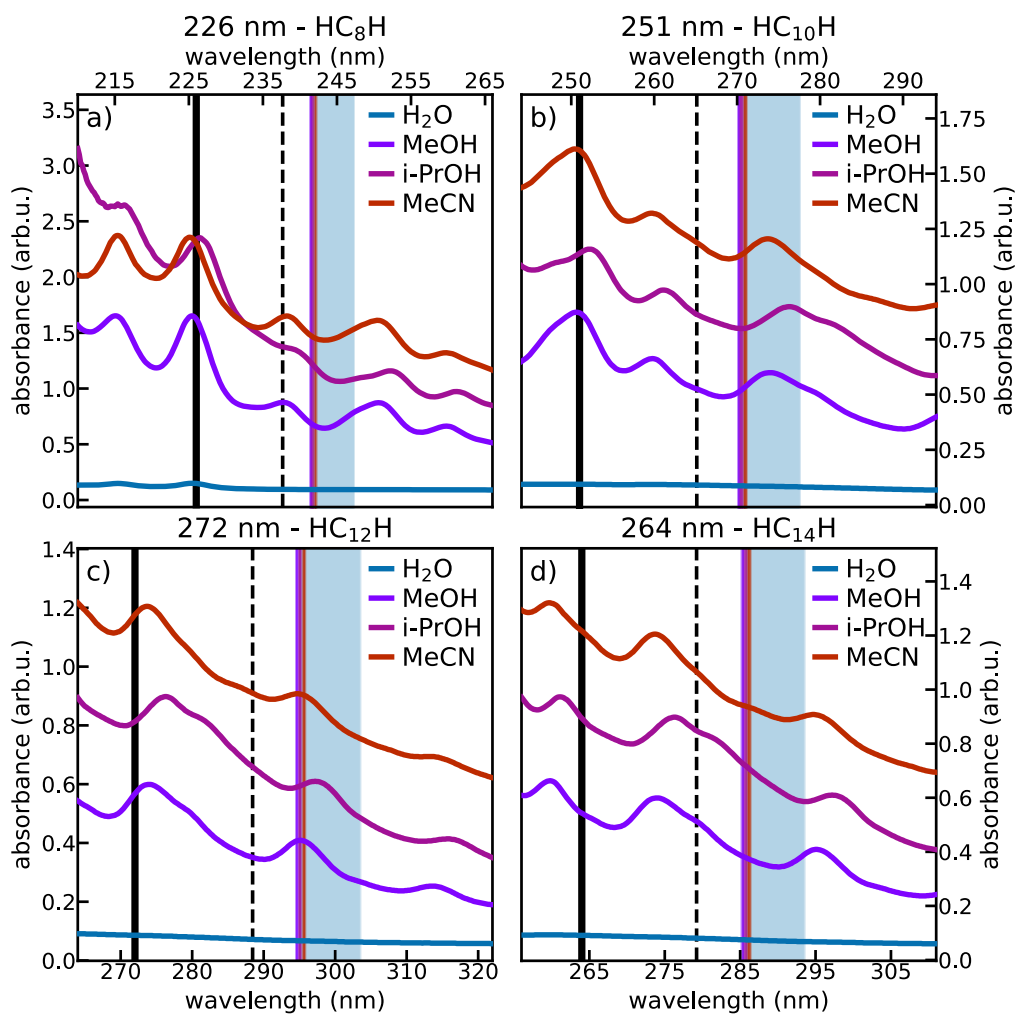

**Figure S8 UV Raman excitation and Raman backscatter photon wavelengths compare to UV-Vis absorption spectra of polyynes mixtures in different solvents.** UV-Vis absorption spectra of water ($H_2O$), methanol (MeOH), isopropanol (i-PrOH), and acetonitrile (MeCN) after 15 minutes of ablations of a graphite target in the same conditions of *in situ* experiments. We diluted the mixtures 10 times with pure solvents to reduce the absorbance and avoid instrument saturation. The thick solid black lines represent the Raman excitation at 226 nm (panel a), 251 nm (panel b), 272 nm (panel c), and 264 nm (panel d). The dashed black lines report the wavelength of the Raman photons of the α mode of $HC_8H$ (panel a), $HC_{10}H$ (panel b), $HC_{12}H$ (panel c), and $HC_{14}H$ (panel d). The violet boxes represent the wavelength range of the OH stretching of water excited at the different Raman excitation wavelengths. The colored lines display the wavelength of the Raman photons of the CH stretching mode of MeOH, i-PrOH, and MeCN excited at the different Raman excitation wavelengths.

Given the self-absorption (SA) issue encountered in UVRR spectra, we established a method to correct our data from SA. In a first approximation, the area of the solvent Raman peaks ($A_s(t)$, where $t$ is the ablation time) should remain constant during the measurements ($A_s(t) = A_s(0)$). Indeed, if the focal conditions and the power deposited are not changing, we expect the same integrated solvent signal at any ablation time. Thus, these Raman features are good quantifiers for the mixture's SA, and we choose them as internal references to correct *in situ* UVRR data from SA. In particular, we select the OH stretching band for water and CH stretching modes at around 3000 cm$^{-1}$ for organic solvents. Even though we could use the CN stretching mode from spectra of acetonitrile ablations, we employed CH ones to keep internal coherence among the organic solvents, helping the comparison and avoiding any under- or over-estimation compared to the other solvents.

In such a way, we can correct the integrated polyynes' signal ($A_p(t)$) from the mixture's SA and precisely evaluate its real dynamics ($A'_p(t)$) as a function of the ablation time ($t$) with the following equation:

$$A'_p(t) = A_p(t)\frac{A_s(0)}{A_s(t)}$$

Beyond the current model, we could gain more precise results by using analytical expressions to calculate $A'_p(t)$, like the ones derived in the work of Hong and Asher.[1] These equations require several relevant physical quantities, like the resonance Raman cross section of H-capped polyynes and byproducts and their molar extinction coefficients, which are not available in the literature and are hardly measurable for these compounds. Moreover, they involve a precise knowledge of experimental parameters, like the area of the synchrotron beam in the focal point and the optical path length in the mixture, which cannot be estimated in our setup with the necessary accuracy. Thus, these fine models are practically unusable in complex environments like our *in situ* measurement scheme. This discussion gives more relevance to our empirical model that can easily be adapted to similar experiments. Indeed, we employed this model in a previous work concerning *in situ* surface-enhanced Raman scattering measurements during ablation [2].

## S3 Conversion model: from integrated UVRR signal to polyynes' concentration

We developed a conversion model to transform UVRR integrated signals associated with the α modes of polyynes, obtained through *in situ* UVRR measurements, to the corresponding polyynes' concentrations. This model serves as a pivotal aspect of our analysis as it establishes a direct link to the production of polyynes.

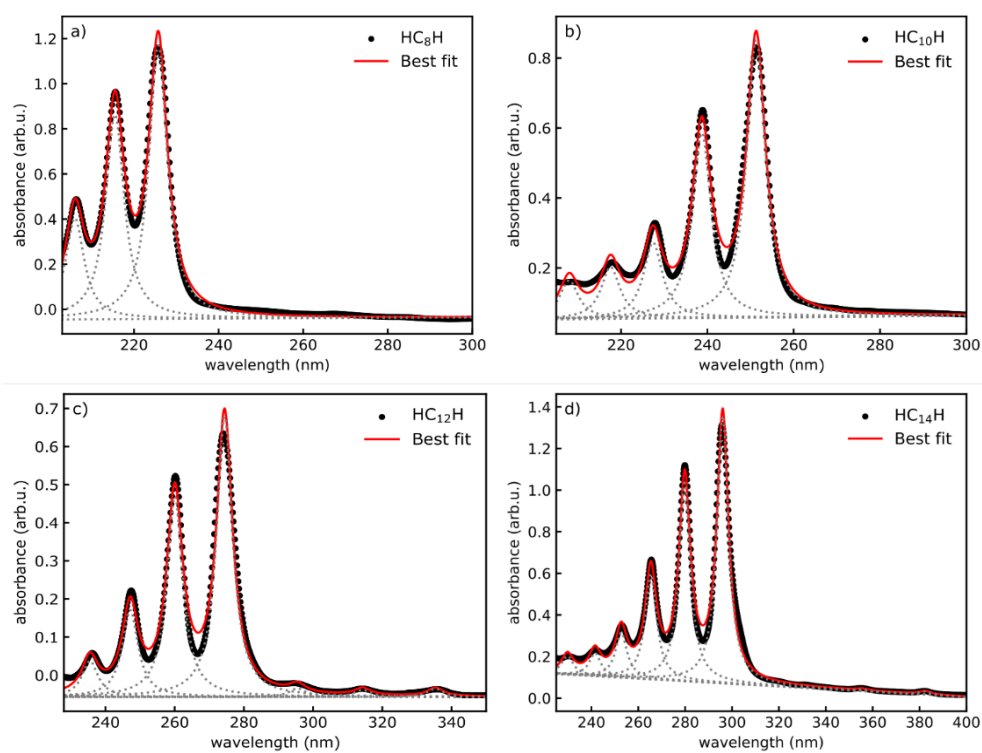

**Figure S9 UV-Vis spectra of H-capped polyynes.** Experimental UV-Vis absorption spectra of size-selected hydrogen-capped polyynes (black dots), namely a) $HC_8H$, b) $HC_{10}H$, c) $HC_{12}H$, and d) $HC_{14}H$. Red solid lines represent best fit functions composed by multi-Lorentzian curves reported as gray dotted lines.

In the initial phase, we collected size-selected samples of $HC_8H$, $HC_{10}H$, $HC_{12}H$, and $HC_{14}H$ from a high-performance liquid chromatography (HPLC) system using the method described in our previous works.[3,4] These chains are dissolved in acetonitrile/water solutions with different proportions (see the caption of

Table S3), corresponding to the mobile phase of our HPLC system. Thus, we conducted UV-Vis absorption measurements on these samples, illustrated in Figure S9. Using these measurements, we employed a multi-Lorentzian fit function, adjusted with a linear baseline, to determine the absorbance of the most prominent peak, corresponding to the 0–0 vibronic transition of each chain. This information allowed us to calculate the concentration of each polyyne, employing Lambert-Beer's law and referring to the extinction molar coefficients outlined in Ref. [5]. The concentration of each size-selected polyyne is reported in Table S3.

| Polyyne | Concentration [$10^{-6}$ mol/L] | Excitation wavelength [nm] | Power on the sample [µW] | Aperture slits [µm] | Acquisition time [s] | CN stretching mode area (pristine solution) |
|---|---|---|---|---|---|---|
| $HC_8H$ | 6.8 ± 0.03 | 226 | 10.7 | 50 | 10 | 36114 ± 51 |
| $HC_{10}H$ | 3.5 ± 0.03 | 251 | 14.3 | 50 | 10 | 21755 ± 22 |
| $HC_{12}H$ | 2.5 ± 0.02 | 272 | 9.4 | 30 | 10 | 1381 ± 2 |
| $HC_{14}H$ | 4.3 ± 0.02 | 264 | 9.2 | 30 | 10 | 2649 ± 4 |

**Table S3** Sample concentrations and UVRR parameters employed to establish the conversion factor for each polyyne. CN stretching mode's UVRR integrated signal is extracted from spectra of pristine solutions. Since the solutions are a mixture of acetonitrile and water, we multiplied the area by ≈1.45 (275/175 MeCN/$H_2O$), ≈1.29 (310/90 MeCN/$H_2O$), ≈1.11 (360/40 MeCN/$H_2O$), and ≈1.06 (372/28 MeCN/$H_2O$) for $HC_8H$, $HC_{10}H$, $HC_{12}H$, and $HC_{14}H$, respectively. The multiplied area is reported in the last column.

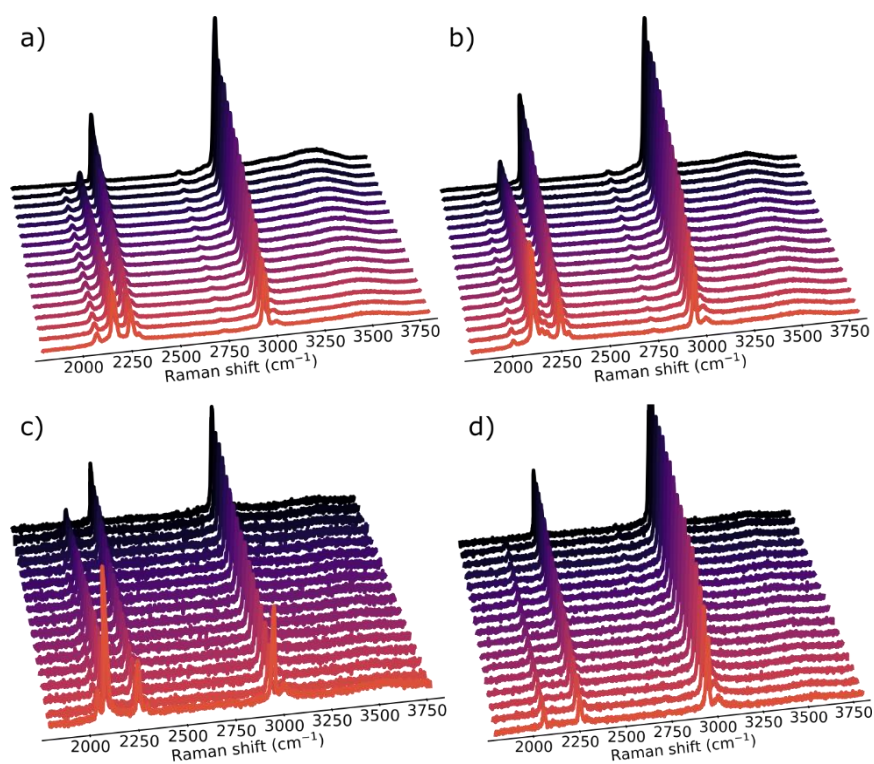

**Figure S10 UVRR spectra of H-capped polyynes at different dilutions.** UV resonance Raman spectra of size-selected hydrogen capped polyynes, namely a) $HC_8H$, b) $HC_{10}H$, c) $HC_{12}H$, and d) $HC_{14}H$ at increased dilution going from orange (maximum concentration) to purple to black (pure solvent).

Following this approach, we systematically obtained UVRR spectra exciting each polyyne at its 0–0 vibronic transition, namely 226, 251, and 272 nm for $HC_8H$, $HC_{10}H$, and $HC_{12}H$. Due to the limitation of the explorable excitation range related to the technical characteristics of the IUVS beamline discussed in the main text, we employed the 0–2 vibronic transition at 264 nm to obtain UVRR spectra of $HC_{14}H$. We collected these spectra using the same parameters of *in situ* UVRR measurements, even if the synchrotron-based radiation power and the collection efficiency varied between these two sets (*i.e.*, *in situ* UVRR measurements and UVRR spectra in Figure S10) of experiments, as listed in Table S3. We measured the pristine polyynes' solutions and collected UVRR spectra after gradually diluting each solution by incrementally adding 20 or 40 µL of a pristine acetonitrile/water solution, maintaining the same proportions as those in the samples containing polyynes. The UVRR spectra of the pristine and diluted solutions for each chain and the reference acetonitrile/water mixtures are shown in Figure S10. We employed a multi-Lorentzian fit function coupled with linear baseline correction to evaluate the UVRR integrated signals of polyynes' α mode in Figure S10, along with that of solvent (MeCN and water) characteristic signals. The values of the corresponding areas are reported in Figure S11a, c, e, and g. To mitigate the distortion introduced by polyynes' self-absorption (SA), similarly to what was discussed in the main text, we used the CN stretching mode of MeCN as an internal reference.

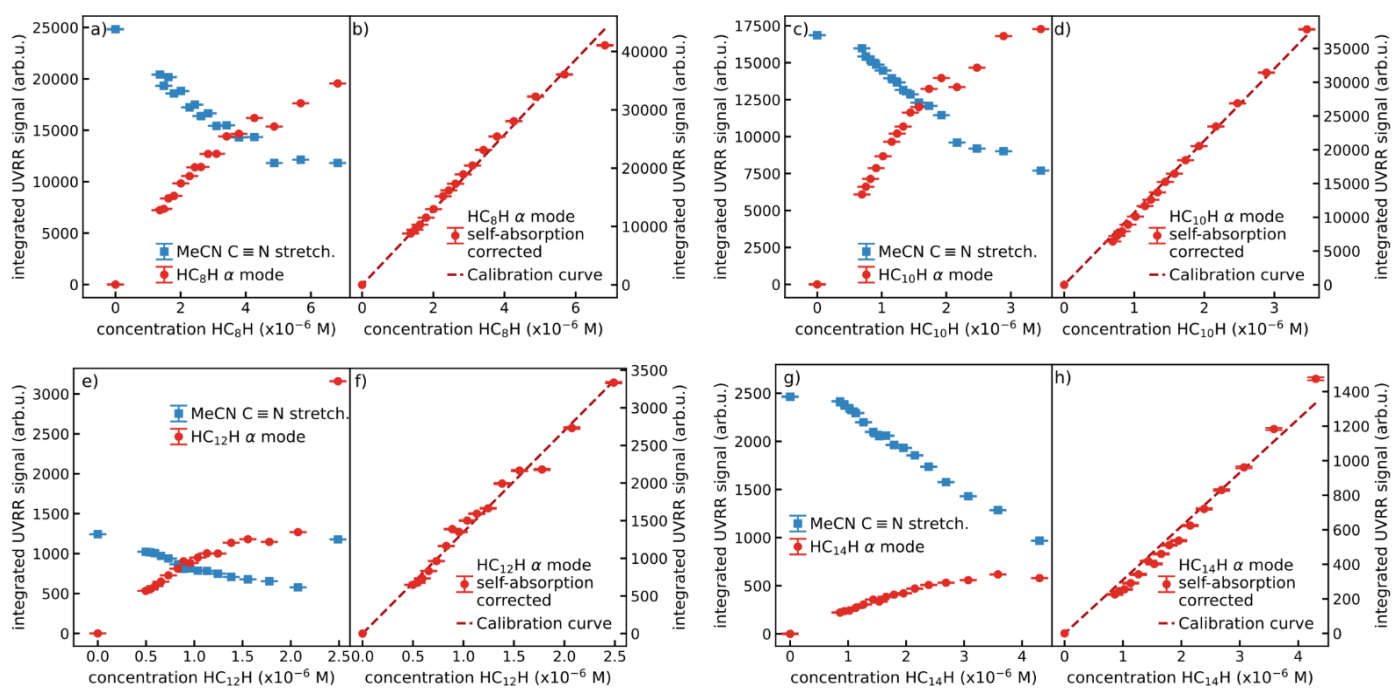

**Figure S11 Polyynes and solvents UVRR integrated signal vs. concentration, uncorrected and corrected from self-absorption.** a) UVRR integrated signal of $HC_8H$'s α mode (red dots) and MeCN CN stretching mode (blue squares) excited at 226 nm as a function of $HC_8H$'s concentration. b) UVRR integrated signal of $HC_8H$'s α mode corrected from self-absorption (red dots) as a function of its concentration. The calibration curve is reported (dark red dashed line). c) UVRR integrated signal of $HC_{10}H$'s α mode (red dots) and MeCN CN stretching mode (blue squares) excited at 251 nm as a function of $HC_{10}H$'s concentration. d) UVRR integrated signal of $HC_{10}H$'s α mode corrected from self-absorption (red dots) as a function of its concentration. The calibration curve is reported (dark red dashed line). e) UVRR integrated signal of $HC_{12}H$'s α mode (red dots) and MeCN CN stretching mode (blue squares) excited at 272 nm as a function of $HC_{12}H$'s concentration. f) UVRR integrated signal of $HC_{12}H$'s α mode corrected from self-absorption (red dots) as a function of its concentration. The calibration curve is reported (dark red dashed line). g) UVRR integrated signal of $HC_{14}H$'s α mode (red dots) and MeCN CN stretching mode (blue squares) excited at 264 nm as a function of $HC_{14}H$'s concentration. h) UVRR integrated signal of $HC_{14}H$'s α mode corrected from self-absorption (red dots) as a function of its concentration. The calibration curve is reported (dark red dashed line).

Consequently, the corrected UVRR area of the α mode of each polyyne ($A'_p(c)$) at each concentration ($c$) is calculated using a modified version of Eq. 1 in the main text, where we substituted the ablation time $t$ with the concentration $c$

$$A'_p(c) = A_p(c)\frac{A_s(0)}{A_s(c)}$$

Here, $A_p(c)$ is the UVRR area of the α mode at polyyne's concentration $c$, $A_s(0)$ and $A_s(c)$ are the UVRR integrated signal of the reference solvent's Raman band, the CN stretching mode of MeCN, in the pristine acetonitrile/water solution ($c = 0$) and at polyyne's concentration $c$. The SA-corrected data are reported in Figure S11b, d, f, and h.

Polyynes-corrected UVRR integrated signals are linearly proportional to their concentration. Using a linear fitting function and forcing the intercept to zero (no polyynes means zero UVRR area), it is possible to find a conversion factor ($k_{UVRR \to c}(p, \lambda)$) to extract the concentration of each polyyne ($p$) at specific Raman excitation wavelengths ($\lambda$). This factor is further adjusted considering all the differences between the *in situ* UVRR data (see Table S1) and this series of spectra (see Table S3). In particular, we multiplied the conversion factor by the ratio between the CN stretching mode's areas of the two experiments, *i.e.*, *in situ* over *ex situ*. In this way, we will achieve the concentration of polyynes during ablation starting from *in situ* UVRR data. The conversion factors are $4.8 \pm 0.05 \cdot 10^9$ (mol/L)$^{-1}$ for $HC_8H$, $6.9 \pm 0.06 \cdot 10^9$ (mol/L)$^{-1}$ for $HC_{10}H$, $3.4 \pm 0.05 \cdot 10^{10}$ (mol/L)$^{-1}$ for $HC_{12}H$, and $5.5 \pm 0.1 \cdot 10^9$ (mol/L)$^{-1}$ for $HC_{14}H$. The errors here reported derive from the fitting procedures used to extract the integrated UVRR signals in *in situ* UVRR, pristine, and diluted spectra and to calculate the conversion factor from the linear calibration curve in Figure S11.

# S4 Linear production rates and saturated growth regime

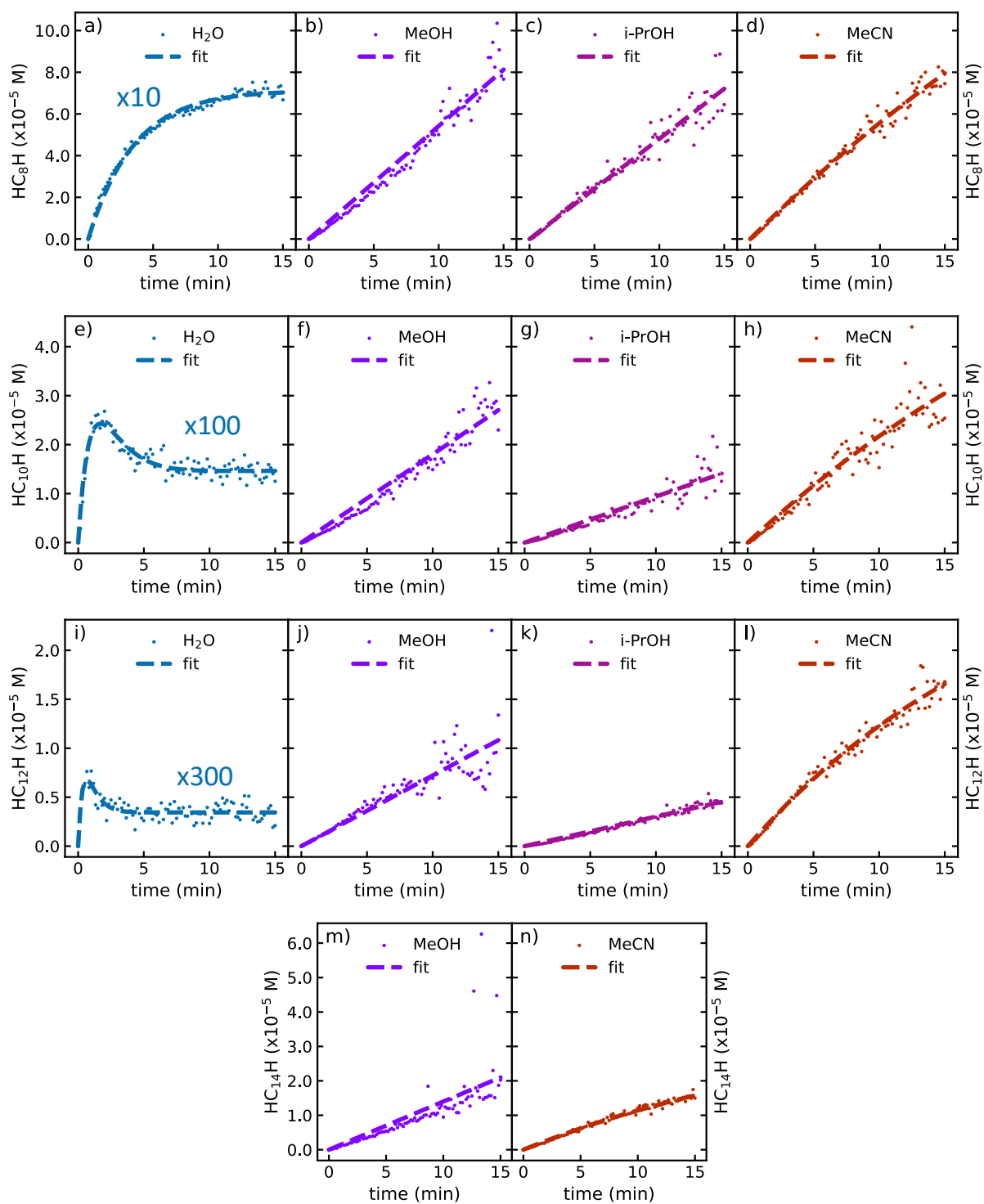

**Figure S12 Evolution and modeling of polyynes concentration during PLAL.** Evolution (colored circles and error bars) and fit (colored dashed lines) result of the SA-corrected concentration of size-selected H-capped polyynes during PLAL in different solvents as a function of the ablation time (refer to Figure 3 for the experimental data). SA-corrected concentration behavior and fit function of $HC_8H$ extracted from *in situ* UVRR data collected at 226 nm in a) water (modeled with Eq. 3 from the main text), b) methanol (Eq. 5), c) isopropanol (Eq. 5), and d) acetonitrile (Eq. 3). SA-corrected concentration behavior and fit function of $HC_{10}H$ extracted from *in situ* UVRR data collected at 251 nm in e) water (Eq. 4), f) methanol (Eq. 5), g) isopropanol (Eq. 5), and h) acetonitrile (Eq. 3). SA-corrected concentration behavior and fit function of $HC_{12}H$ extracted from *in situ* UVRR data collected at 272 nm in i) water (Eq. 4), j) methanol (Eq. 5), k) isopropanol (Eq. 5), and l) acetonitrile (Eq. 3). SA-corrected concentration behavior and fit function of $HC_{14}H$ extracted from *in situ* UVRR data collected at 264 nm in m) methanol (Eq. 5) and n) acetonitrile (Eq. 3).

|  | HC$_8$H [·10$^{-6}$ M/min] | HC$_{10}$H [·10$^{-6}$ M/min] | HC$_{12}$H [·10$^{-6}$ M/min] | HC$_{14}$H [·10$^{-6}$ M/min] |
|---|---|---|---|---|
| H$_2$O | 2.04 ± 0.05 | – | – | – |
| MeOH | 5.42 ± 0.07 | 1.80 ± 0.03 | 0.72 ± 0.02 | 1.40 ± 0.08 |
| i-PrOH | 4.81 ± 0.06 | 0.94 ± 0.02 | 0.302 ± 0.003 | – |
| MeCN | 6.22 ± 1.95 | 2.49 ± 1.58 | 1.54 ± 0.26 | 1.35 ± 0.24 |

**Table S4** Production rates in M/min (M is molarity, mol/L) extracted from the fitting procedures of the evolution of the corresponding concentration curves (see Figure S12). See the main text for further details. The fit errors are reported as well.

|  |  | HC$_8$H | HC$_{10}$H | HC$_{12}$H | HC$_{14}$H |
|---|---|---|---|---|---|
| H$_2$O | Saturation concentration [·10$^{-6}$ M] | 7.1 ± 0.1 | 0.15 ± 0.02 | 0.011 ± 0.005 | – |
|  | Saturation time [min] | 10.5 ± 0.2 | 0.44 (≈26 s) ± 0.44 | 0.14 (≈8 s) ± 0.06 | – |
| MeCN | Saturation concentration [·10$^{-4}$ M] | 2.8 ± 0.6 | 0.9 ± 0.4 | 0.33 ± 0.03 | 0.39 ± 0.04 |
|  | Saturation time [min] | 137 ± 32 | 106 ± 51 | 64 ± 8 | 87 ± 12 |

**Table S5** Saturation concentration (in M) and saturation time to reach 95 % of the saturation concentration (in min) extracted from the fitting procedures of the evolution of the concentration curves in water and acetonitrile (see Figure S12). The fit errors are reported as well.

| HC$_{16}$H | HC$_{18}$H | HC$_{20}$H | HC$_{22}$H | HC$_{24}$H | HC$_{26}$H | HC$_{28}$H | HC$_{30}$H |
|---|---|---|---|---|---|---|---|
| 3.99 · 10$^{-6}$ | 1.37 · 10$^{-6}$ | 4.74 · 10$^{-7}$ | 1.63 · 10$^{-7}$ | 5.63 · 10$^{-8}$ | 1.94 · 10$^{-8}$ | 6.69 · 10$^{-9}$ | 2.31 · 10$^{-9}$ |

**Table S6** Saturation concentrations (in M) extracted from fitting the data in Figure 5a in the main text (or in Table S5).